\documentclass[preprint,floats,aps,epsfig,nofootinbib,11pt]{revtex4}

\evensidemargin=\oddsidemargin
\usepackage{graphicx}
\usepackage{bm}

\begin{document}

\renewcommand{\baselinestretch}{1.2}
\def\fbi{\rm fb^{-1}}
\def\mev{\rm MeV}
\def\gev{\rm GeV}
\def\tev{\rm TeV}
\def\be{\begin{equation}}
\def\ee{\end{equation}}
\def\bea{\begin{eqnarray}}
\def\eea{\end{eqnarray}}
\def\lsim{\mathrel{\raise.3ex\hbox{$<$\kern-.75em\lower1ex\hbox{$\sim$}}}}
\def\gsim{\mathrel{\raise.3ex\hbox{$>$\kern-.75em\lower1ex\hbox{$\sim$}}}}

\def\del{\partial }

\preprint{
 {\vbox{
\hbox{TUHEP-TH-04149}
 \hbox{MADPH--05--1431}
}}}

\vspace*{1cm}

\title{\Large Anomalous gauge couplings of the Higgs boson\\ at high energy photon colliders}
\bigskip
\author{Tao Han$^{1,3,4}$, Yu-Ping Kuang$^{2,3}$, and Bin Zhang$^{3,1}$ }
\address{$^1$Department of Physics, University of Wisconsin, Madison, WI 53706,
U.S.A.\footnote{Email address: than@physics.wis.edu.}\\
$^2$CCAST (World Laboratory), P.O. Box 8730, Beijing 100080, P.R. China\\
$^3$Center for High Energy Physics, Tsinghua University, Beijing 100084, P.R.
China\footnote{Mailing address for Yu-Ping Kuang and Bin Zhang. Email:
ypkuang@mail.tsinghua.edu.cn,
zb@mail.tsinghua.edu.cn. }\\
$^4$Institute of Theoretical Physics, Academia Sinica, Beijing 100080, P.R. China}

\begin{abstract}

We study the sensitivity of testing the anomalous gauge couplings $g_{HVV}$'s of the
Higgs boson in the formulation of linearly realized gauge symmetry  via the processes
$\gamma\gamma\to ZZ$ and $\gamma\gamma\to WWWW$ at polarized and unpolarized photon colliders
based on $e^+e^-$ linear colliders of c.m.~energies
500 GeV, 1 TeV, and 3 TeV. Signals beyond the standard model (SM)
and SM backgrounds are carefully studied.
We propose certain kinematic cuts to suppress the standard model backgrounds.
For an integrated  luminosity of 1 ab$^{-1}$, we show that (a) $\gamma\gamma\to ZZ$ can
provide a test of $g_{H\gamma\gamma}$ to the $3\sigma$ sensitivity of $O(10^{-3}\--10^{-2})$
TeV$^{-1}$ at a 500 GeV ILC, and
$O(10^{-3})$ TeV$^{-1}$ at a 1 TeV ILC and a
3 TeV CLIC, and (b) $\gamma\gamma\to WWWW$
at a 3 TeV CLIC can test all the anomalous couplings $g_{HVV}$'s to the $3\sigma$ sensitivity of
$O(10^{-3}\--10^{-2})$ TeV$^{-1}$.

\null\noindent PACS number(s): 12.60.Fr, 13.88.+e, 14.80.Cp
\end{abstract}
\maketitle

\bigskip

\section{INTRODUCTION}

Probing the  mechanism  of electroweak symmetry breaking (EWSB)
is one of the most important tasks at TeV-scale colliders.
The direct search in the CERN LEP experiments
for the Higgs boson, which is related to
EWSB and mass generation as predicted in the standard model (SM),
 sets a lower bound on its mass $m_H>114.4$ GeV \cite{PDG}. The
precision electroweak data  favors a light Higgs boson with
a mass $m_H\le 186$ GeV at  95\%~C.L. \cite{MH,PDG}.
It has been shown that  convincing
evidence for a SM-like Higgs boson would be discovered at the
CERN Large Hadron Collider (LHC) \cite{ATLAS-CMS}  and be studied in
great detail at $e^+e^-$ International linear colliders (ILC) \cite{ILC}.
Once a Higgs boson is found at TeV-scale colliders, it is of fundamental
importance to check if the Higgs boson is SM-like by studying
its couplings to the SM particles.
Primarily due to the ``naturalness" argument  \cite{unnaturalness}
for a light Higgs boson, it is widely believed that  physics beyond the SM
must exist  at a scale $\Lambda$ near the order of TeV.
In particular, if there are no new light particles observed
other than the Higgs boson in the next generation collider experiments,
it is even more pressing to determine the Higgs boson
couplings as accurately as possible to seek for hints for new physics
beyond the SM.

To extend the structure of the SM in a model-independent approach,
it is customary to formulate  new physics effects by linearly realizing
the gauge symmetry \cite{linear,anomal}.
After integrating out heavy degrees of freedom at the scale $\Lambda$,
the leading effects at low energies can be parameterized by
the effective interactions
\begin{eqnarray}                              
{\cal L}_{\mbox{eff}} ~\,=~\, \sum_n \frac{f_n}{\Lambda^2} {\cal O}_n \,,
\label{eff}
\end{eqnarray}
where $f_n$'s are dimensionless ``anomalous couplings",
and ${\cal O}_n$ the gauge-invariant  dimension-6 operators, constructed
from the SM fields. If $\Lambda$ appropriately parameterizes the new
physics scale (such as the mass of the next resonance), then one would
expect $f_n$'s to be of the order of unity.
The anomalous couplings of the Higgs boson and gauge bosons are of special
interest since they may be directly related to the mechanism of EWSB.
Theoretical studies of testing the anomalous gauge couplings of the Higgs boson
already exist in the literature
for the LHC \cite{Eboli,deCampos,Zeppenfeld,HKYZ03,ZKHY03},
and for the ILC \cite{HZZ,BHLMZ,G-G}.
In this paper, we extend the literature  by performing a systematic study of testing the
anomalous gauge couplings of the Higgs boson
at polarized and unpolarized photon colliders based on $e^+e^-$ linear colliders
 of various energies. We consider the processes
 \begin{eqnarray}
 \gamma\gamma \to ZZ,\quad \gamma\gamma \to WWWW.\nonumber
 \end{eqnarray}
 We show that a photon collider can be more beneficial,
 and the sensitivity to the couplings can be improved over the current results
 from other colliders.

This paper is organized as follows. In Sec.~\ref{anom}, we review the operators
for the anomalous gauge couplings of the Higgs boson and present the
current constraints on them. In Sec.~\ref{photon}, we summarize the backscattered
photon spectrum adopted in our calculations, and give a brief sketch for calculating the cross
sections at the polarized and unpolarized photon colliders based on the $e^+e^-$ linear colliders.
Secs.~\ref{zz}  and \ref{wwww} are the studies on the
$\gamma\gamma\to ZZ$ and $\gamma\gamma\to WWWW$ processes,
respectively. The summary of the results is given in Sec.~\ref{sum}.

\section{Anomalous gauge couplings of the Higgs boson}
\label{anom}

In the formulation of linearly realized gauge symmetry,
the $C$ and $P$ conserving dimension-6 effective operators of our current
interests involving the $SU(2)$ gauge field $W^i_\mu$,  the $U(1)$
gauge field $B_\mu$ as well as a Higgs doublet $\Phi$
are given by  \cite{linear,anomal,G-G}
\begin{eqnarray}                    
&& {\cal O}_{BW} =  \Phi^{\dagger} \hat{B}_{\mu \nu}
\hat{W}^{\mu \nu} \Phi ,  \qquad \qquad\qquad
{\cal O}_{\Phi,1} = \left ( D_\mu \Phi \right)^\dagger
\Phi^\dagger \Phi \left ( D^\mu \Phi \right ),   \label{Obliq}\\
&&{\cal O}_{\Phi,2} =\frac{1}{2}
\partial^\mu\left ( \Phi^\dagger \Phi \right)
\partial_\mu\left ( \Phi^\dagger \Phi \right),\quad\quad~
{\cal O}_{\Phi,3} =\frac{1}{3} \left(\Phi^\dagger \Phi \right)^3, \label{self}\\
&&
{\cal O}_W  = (D_{\mu} \Phi)^{\dagger}\hat{W}^{\mu \nu}  (D_{\nu} \Phi),
\qquad\qquad
{\cal O}_B  =  (D_{\mu} \Phi)^{\dagger} \hat{B}^{\mu \nu}(D_{\nu} \Phi),
\nonumber\\
&&
{\cal O}_{WW} = \Phi^{\dagger} \hat{W}_{\mu \nu}
\hat{W}^{\mu \nu} \Phi , \qquad\qquad\quad~
{\cal O}_{BB} = \Phi^{\dagger} \hat{B}_{\mu \nu} \hat{B}^{\mu\nu} \Phi,
\label{O}
\end{eqnarray}
where $\hat B_{\mu\nu}$ and $\hat W_{\mu\nu}$ stand for
\begin{eqnarray}
\hat{B}_{\mu \nu} = i \frac{g'}{2} B_{\mu \nu},\qquad
\hat{W}_{\mu \nu} = i \frac{g}{2} \sigma^a W^a_{\mu \nu},\nonumber
\end{eqnarray}
in which $g$ and $g^\prime$ are the $SU(2)$ and $U(1)$ gauge
couplings, respectively.

Precision electroweak data and the measurements of the triple-gauge-boson couplings
give considerable constraints on some of the anomalous couplings $f_n/\Lambda^2$
in Eq.~(\ref{eff}) \cite{ZKHY03,BHLMZ,G-G}. For instance, the oblique correction parameters $S$ and
$T$ \cite{peskin} give rise to rather stringent constraints on the anomalous coupling constants
$f_{BW}$ and $f_{\Phi,1}$ in Eq.~(\ref{Obliq}) \cite{G-G,ZKHY03}. The $1\sigma$ and $2\sigma$
contours for $f_{BW}$ and $f_{\Phi,1}$ from the updated experimental values of $S$ and $T$ are given
in Ref.~\cite{ZKHY03}. Assuming either $f_{BW}$ or  $f_{\Phi,1}$ dominance, the $2\sigma$
constraints obtained are quite strong \cite{ZKHY03},
\begin{eqnarray}
-0.07 <  {f_{BW}\over (\Lambda/{\tev})^2}  < 0.04, \quad
-0.02 <  {f_{\Phi,1}\over (\Lambda/{\tev})^2}  < 0.02.\nonumber
\end{eqnarray}
The next two operators in Eq.~(\ref{self}) are purely Higgs boson  self-interactions,
and lead to corrections to the Higgs triple and quartic vertices. They
have been dedicatedly studied in \cite{BHLMZ}  at linear colliders
and we will not pursue them further.
However, the present experimental observables are not sensitive to the four anomalous coupling
operators $O_W,~O_{WW},~O_B$ and $O_{BB}$ (with anomalous coupling constants
$f_W/\Lambda^2,~f_{WW/\Lambda^2},~f_B/\Lambda^2$ and $f_{BB}/\Lambda^2$) in Eq.~(\ref{O}). The
constraints from the existing experiments and the requirement of unitarity of the $S$ matrix element
on these four anomalous coupling constants are rather weak.
We summarize the above 
constraints on those four anomalous couplings in TABLE~\ref{summary2}.
The results are obtained by assuming
only one anomalous coupling exists each time.
\begin{table}[h]
\caption{Current $2\sigma$ constraints on $f_n/\Lambda^2$
from existing studies.
The results are obtained by assuming
only one anomalous coupling exists each time.}
\vspace*{2mm}
\renewcommand{\baselinestretch}{1}
\tabcolsep 45pt
\begin{tabular}{lc}
\hline\hline
Constraints from & $f_n/\Lambda^2$  in TeV$^{-2}$ \\
\hline
 Precision EW fit \cite{ZKHY03}:
 & $-6\leq\frac{f_W}{\Lambda^2}\leq 5$   \\ \\
 & $4.2\leq\frac{f_B}{\Lambda^2}\leq2.0$ \\ \\
 &$-5.0\leq\frac{f_{WW}}{\Lambda^2}\leq 5.6$ \\ \\
  &$17\leq\frac{f_{BB}}{\Lambda^2}\leq 20$ \\ \\ \\
Triple gauge coupling \cite{G-G}
& $-31\leq\frac{f_W+f_B}{\Lambda^2}\leq 68$ \\ \\
LEP2 Higgs searches \cite{Eboli}:
& $-7.5\leq \frac{f_{WW(BB)}}{\Lambda^2}\leq18 $ \\ \\
Unitarity (at $\sqrt{s}$=2 TeV) \cite{Gounaris}:  &$|\frac{f_B}{\Lambda^2}|\leq
24.5;\quad |\frac{f_W}{\Lambda^2}|\leq 7.8$ \\\\
 &$-160\leq |\frac{f_{BB}}{\Lambda^2}|\leq 197;\quad |\frac{f_{WW}}{\Lambda^2}|\leq 39.2$ \\ \\
 \hline\hline
\label{summary2}
\end{tabular}
\end{table}

It is perhaps more intuitive to express the new operators in terms of  couplings
of the explicit physical component fields.
Taking into account the mixing between $W^3_\mu$ and $B_\mu$,
the effective couplings of the Higgs boson $H$ and the electroweak gauge bosons $V$
($V=\gamma,\ W^\pm,\ Z$) in Eqs.~(\ref{eff}) and (\ref{O}) can be cast into
\cite{G-G}
\begin{eqnarray}                         
{\cal L}^H_{\rm eff}&=&g_{H\gamma\gamma}HA_{\mu\nu}A^{\mu\nu}
+g^{(1)}_{HZ\gamma}A_{\mu\nu}Z^\mu\partial^\nu H
+g^{(2)}_{HZ\gamma}HA_{\mu\nu}Z^{\mu\nu}
+g^{(1)}_{HZZ}Z_{\mu\nu}Z^\mu
\partial^\nu H\nonumber\\
&&
+g^{(2)}_{HZZ}HZ_{\mu\nu}Z^{\mu\nu}
+g^{(1)}_{HWW}(W^+_{\mu\nu} W^{-\mu}\partial^\nu H+{\rm h.c.})
+g^{(2)}_{HWW}HW^+_{\mu\nu}W^{-\mu\nu},
\label{LHeff}
\end{eqnarray}
where the anomalous couplings $g_{HVV}$'s (of dimension $-1$)
are related to those Lagrangian parameters $f_n$'s by
\begin{eqnarray}                           
\displaystyle &&g^{}_{H\gamma\gamma}=-\alpha\frac{s^2(f_{BB}
+f_{WW})}{2},\nonumber\\
&& g^{(1)}_{HZ\gamma}=\alpha\frac{s(f_W-f_B)}{2c},
~~~~~~~~~
g^{(2)}_{HZ\gamma}=\alpha\frac{s[s^2f_{BB}
-c^2f_{WW}]}{c},\nonumber\\
&&g^{(1)}_{HZZ}=\alpha\frac{c^2f_W+s^2f_B}{2c^2},
~~~~~~~~
g^{(2)}_{HZZ}=-\alpha\frac{s^4f_{BB}
+c^4f_{WW}}{2c^2},\nonumber\\
&&g^{(1)}_{HWW}=\alpha\frac{f_W}{2},~~~~~~~~~~~~~~~~~~~
g^{(2)}_{HWW}=-\alpha f_{WW}, \label{g}
\end{eqnarray}
with the weak mixing $s\equiv \sin\theta_W,~c\equiv \cos\theta_W$ and
$\alpha=gM_W/\Lambda^2\approx 0.053\ {\rm TeV}^{-1}\approx {1/(19\ \tev)}$ .
Roughly speaking, an order unity coupling of $f_n$ translates to
$g^{(i)}_{HVV} \sim 1/(20$ TeV)=0.05 TeV$^{-1}$.

Since new physics responsible for the mechanism of the EWSB
is more likely to show up with the Higgs couplings to gauge bosons, these couplings
should be tested as thoroughly as possible at future high energy colliders
\cite{Eboli,deCampos,Zeppenfeld,HKYZ03,ZKHY03,HZZ,BHLMZ,G-G}.
At the LHC, it is shown in
 Ref.~\cite{ZKHY03} that the most sensitive constraints on $f_W/\Lambda^2$ and
 $f_{WW}/\Lambda^2$  will be from the measurement of the gauge-boson scattering
 $W^+W^+\to W^+W^+$.
 The $2\sigma$ level constraints obtained on these two anomalous couplings are
\begin{equation}
-1.4~{\rm TeV}^{-2}<f_W/\Lambda^2<1.2~{\rm TeV}^{-2},\quad {\rm and}\quad
2.2~{\rm TeV}^{-2}\le f_{WW}/\Lambda^2<2.2~{\rm TeV}^{-2},
\label{LHCww}
\end{equation}
which may reach the parameter regime sensitive to TeV-scale new physics.
Those processes are insensitive to $f_B/\Lambda^2$ and $f_{BB}/\Lambda^2$
however \cite{ZKHY03}.
 At $e^+e^-$ linear colliders on the other hand, the
 anomalous couplings $g^{(1)}_{HZZ}$ and $g^{(2)}_{HZZ}$
 can be constrained at the $2\sigma$ sensitivity to $(10^{-3}\-- 10^{-2})$ TeV$^{-1}$
 from the Higgs-strahlung process $e^+e^-\to Z^*\to ZH$ \cite{HZZ}.

  We will see in the later sections that, at photon colliders, the sensitivities
to probe those couplings can be improved.

\section{Backscattering photon spectrum and gauge-boson prouction}
\label{photon}

By means of laser backscattering, a photon collider  \cite{lumin,pol-ee}
can be built on an $e^+e^-$ linear collider.
Let $m_e$ and $E_e$ be the incident electron mass and energy, respectively; $\omega_0$ and $\omega$
be the laser photon and the backscattered photon energies, respectively; $\sqrt{s}$
be the  $e^+e^-$ center-of-mass energy; $\sqrt{\hat{s}}$ be the center-of-mass energy of the
backscattered photon; and $M_V$ be the mass of the produced weak gauge boson.
For an unpolarized photon collider, the cross section $\sigma(s)$ for the production of $2n$
weak gauge bosons at the photon collider can be obtained by convoluting the subprocess
cross section $\sigma(\hat{s})$ with the photon luminosity  at an $e^+e^-$ linear collider
\begin{eqnarray}                                
\sigma(s)=\int^{x_{max}}_{2nM_V/\sqrt{\hat{s}}}dz \frac{dL_{\gamma\gamma}}
{d z}\sigma(\hat{s} ),
~~~~\hat{s}=z^2 s,
\label{sigma-unp}
\end{eqnarray}
where
$x_{max}=\omega_{max}/E_e$, and $~dL_{\gamma \gamma}/dz~$ is the photon luminosity defined as
\begin{eqnarray}                                
\frac{dL_{\gamma\gamma}}{d z} = 2z \int^{x_{max}}_{z^2 /x_{max}}
 \frac{dx}{x}F_{\gamma/e}(x) F_{\gamma/e}(z^2/x)\,.
\end{eqnarray}
The energy spectrum $F_{\gamma/e}(x)$ of the backscatterred photon at an unpolarized photon collider
is given by \cite{lumin}
\begin{eqnarray}                         
F_{\gamma/e}(x) &=& \frac{1}{D(\xi)}\bigg[1 - x + \frac{1}{1-x} -\frac{4x}{
\xi(1-x)} + \frac{4x^2}{\xi^2(1-x)^2}\bigg], \label{F} \\
D(\xi)  &=&\left(1 - \frac{4}{\xi}-\frac{8}{\xi^2}\right)\ln(1+\xi) +
\frac{1}{2} + \frac{8}{\xi} - \frac{1}{2(1+\xi)^2}, \label{D}
\end{eqnarray}
where $~\xi = 4E_e\omega_0/ m_e^2$,
and $~x=\omega /E_e $ is the energy fraction carried by the backscattered photon.
$~F_{\gamma/e}(x)~$ vanishes for $~x>x_{max}=\omega_{max}/E_e
=\xi/(1+\xi)~$. In order to avoid the creation of $ e^+ e^- $ pairs by the
interaction of the incident and backscattered photons, we require
$~\omega_0 x_{max}\leq m_e^2/E_e~$ which implies $~\xi \leq 2+2\sqrt{2}
\approx 4.8~$.  For the choice of $~\xi = 4.8~$, we obtain
\begin{eqnarray}                                   
x_{max} \approx 0.83, \quad D(\xi) \approx 1.8 .
\end{eqnarray}
We will take these values for the unpolarized spectrum
in our numerical calculations unless stated otherwise.

\begin{figure}[tb]
\includegraphics[width=12truecm,clip=true]{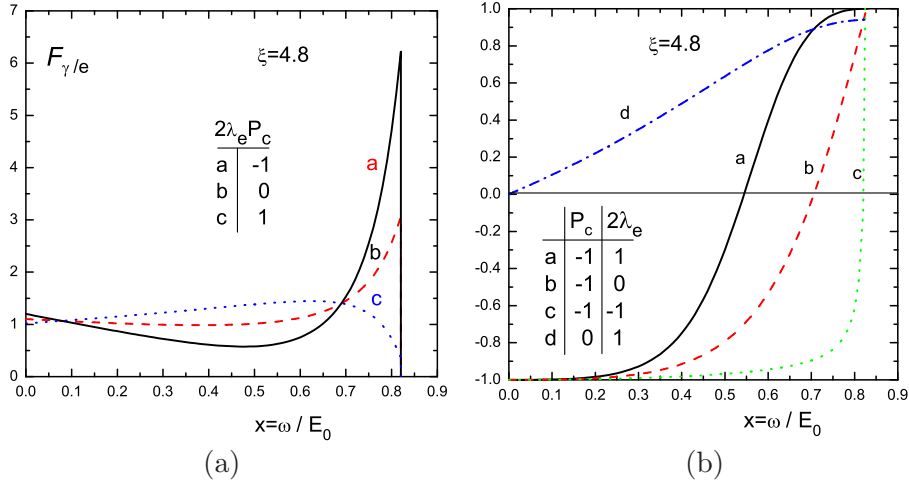}
\null\vspace{-0.3cm} \caption{(a) Photon energy distribution and
(b) averaged helicity distributions with $x=\omega/E_e$  for
various values of $2\lambda_eP_c$ at polarized photon colliders
(based on the formulas given in Ref.~\cite{TESLA6}).}
\label{mean}
\end{figure}
As for a polarized photon collider, let $P_c$ be the polarization of the initial laser,
$\lambda_e$  the polarization of the electron beam,
the energy spectrum of the
backscattered photon beam is
\begin{eqnarray}                      
\nonumber
F_{\gamma/e}(x,\lambda_e,P_c)&=&\frac{1}{D(\xi, \lambda_e, P_c)}\bigg[1-x+\frac{1}{1-x}-
\frac{4x}{\xi(1-x)}+\frac{4x^2}{\xi^2(1-x)^2}\\
&-&2\lambda_eP_c\bigg(\frac{x}{1-x}-\frac{2x^2}{(1-x)^2\xi}\bigg)(2-x)\bigg],
\label{F'} \\
\nonumber
D(\xi, \lambda_e,P_c) & =& (1 - \frac{4}{\xi}-\frac{8}{\xi^2})\ln(1+\xi) + \frac{1}{2}
+ \frac{8}{\xi} - \frac{1}{2(1+\xi)^2}\\
&+&2 \lambda_e P_c \left[ (1+2/\xi)\ln(\xi+1)-5/2+1/(\xi+1)
-\frac{1}{2(\xi+1)^2} \right].
\label{D'}
\end{eqnarray}
It is shown in Ref.~\cite{Ginzburg} that the energy spectrum of the
colliding photons peaks in a narrow region near the high
energy end ($80\%$ of the electron energy) if $2\lambda_e P_c=-1$,
as demonstrated in Fig.~{\ref{mean}}(a) by the solid curve. For comparison,
the dashed curve presents the unpolarized photon spectrum. The choice of
polarizations improves the monochromatization and enhances the effective
energy of the photon collider. Another important measure is the average
polarization of the backscattered photon beam. Figure~{\ref{mean}}(b)
illustrates the average percentage polarization for various choices of
initial beam polarizations. The solid curve is the preferred choice in
terms of the energy spectrum and it is also desirable that it yields
almost purely right-handed polarized beam near the peak $x\sim 0.8$.

After integrating over the azimuthal angles, the differential cross section
can be expressed as  \cite{Ginzburg}
\begin{eqnarray}                      
\label{dN'}
&& d\sigma(\hat s)_{\gamma\gamma\rightarrow nV}
=d\sigma_+ + \zeta_2\tilde{\zeta}_2 d\sigma_-, \\
&& d\sigma_\pm = \frac{1}{4}(|M_{++}|^2+|M_{--}|^2\pm |M_{+-}|^2\pm
|M_{-+}|^2) \ dPS_n \label{amplitude}
\end{eqnarray}
where  $M_{\sigma_1\sigma_2}$ are the helicity amplitudes
for two photons with helicity $\sigma_1,\ \sigma_2$,
and $dPS_n$ the $n$-body phase space element.
$\zeta_2$ and  $\tilde\zeta_2$ are the Stokes parameters \cite{Ginzburg}
for the two colliding photon beams
\begin{eqnarray}
&&\zeta_2=\frac{C_{20}}{C_{00}},\quad  r\equiv\frac{x}{\xi(1-x)}, \nonumber\\
&&C_{00}=\frac{1}{1-x}+1-x-4r(1-r)
-2\lambda_e P_c r \xi(2r-1)(2-x)\nonumber
\\
&&C_{20}=2\lambda_e r
\xi[1+(1-x)(2r-1)^2]
-P_c(2r-1)(\frac{1}{1-x}+1-x),\nonumber
\end{eqnarray}
and $\tilde{\zeta}_2$ is of the same form for the other photon.

We are considering the anomalous gauge couplings with the existence
of a light Higgs boson. For definitiveness, we will take its mass to be in the range
\begin{eqnarray}                            
115~{\rm GeV}\le m_H\le 300~{\rm GeV}.
\end{eqnarray}
When searching for anomalous $HVV$ couplings,
we consider the signal as the excess or deficit from the SM prediction.
The backgrounds are the SM expectation corresponding to $f_n/\Lambda^2=0$.
We thus define the background cross section
$\sigma_B$ and the signal cross section $\sigma_S$ by
\begin{eqnarray}                         
\sigma_B\equiv \sigma(f_n/\Lambda^2=0),\qquad
\sigma_S\equiv \sigma(f_n/\Lambda^2\ne 0)-\sigma_B.
\label{sigmaBsigmaS}
\end{eqnarray}
With the corresponding background and signal event numbers $N_B$
and $N_S$,  the statistical significance $\sigma_{stat}$ is defined by
\begin{eqnarray}                         
\sigma_{stat}\equiv \frac{N_S}{\sqrt{N_S+N_B}}.
\label{N}
\end{eqnarray}

In the following numerical calculations,
we take the  integrated luminosity of an $e^+e^-$ collider to be
$\int{\cal L}dt=1 ~{\rm ab}^{-1}$,
which corresponds to about a two-year run at 0.5 TeV and 1 TeV for the International
Linear Collider (ILC) roughly estimated from the TESLA Technical Design Report \cite{TESLA1}, and
about a 1-year run at 3 TeV for the CERN Compact Linear Collder (CLIC) \cite{CLIC}.

\section{Sensitivity to the anomalous couplings from $\bm{\gamma\gamma\to ZZ}$
\label{zz} }

When considering gauge boson pair production at a photon collider to probe the EWSB
beyond the SM,  as  pointed out in Ref.~\cite{BC},
the $\gamma\gamma\to W^+W^-$ process suffers from a large
tree-level SM background, while $\gamma\gamma\to ZZ$ is free from
tree-level SM backgrounds. We therefore concentrate on this process in this
section.

\begin{figure}[h]
\includegraphics[width=5truecm,clip=true]{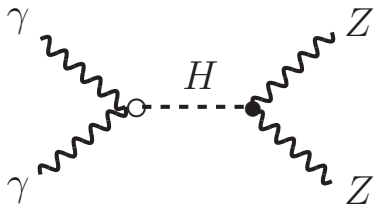}
\caption{Feynman diagram for the signal process $\gamma\gamma\to ZZ$. The vertex with
a circle $\circ$ stands for the anomalous $H\gamma\gamma$ interaction
with the anomalous coupling constant $g_{H\gamma\gamma}$,
and the vertex with a black dot {\small$\bullet$} contains the SM coupling
and the anomalous $HZZ$ interactions with the anomalous couplings
$g^{(1)}_{HZZ}$ and $g^{(2)}_{HZZ}$.}
\label{phph-ZZ}
\end{figure}

With the anomalous $H\gamma\gamma$ coupling, the signal process $\gamma\gamma\to ZZ$ can formally have
a tree level contribution shown in Fig.~\ref{phph-ZZ},
in which the vertex with a circle contains only the anomalous $H\gamma\gamma$
interaction with the coupling $g_{H\gamma\gamma}$, and the vertex with a black dot
 contains the SM  interaction as well as  the anomalous $HZZ$ interactions
 with the anomalous coupling $g^{(1)}_{HZZ}$ and $g^{(2)}_{HZZ}$.
 Our calculation shows that, for reasonable values of $g^{(1)}_{HZZ}$ and $g^{(2)}_{HZZ}$
 and center of mass energy, the contribution of the
anomalous $HZZ$ interactions to the cross section is only a few percent relative to the
SM contribution. Therefore the process $\gamma\gamma\to ZZ$
mainly tests the anomalous couplings $g_{H\gamma\gamma}$ which is
related to $(f_{BB}+f_{WW})/\Lambda^2$ [cf.~Eq.~(\ref{g})].
Studying $W^+W^+$ scattering at the LHC on the other hand
provides a sensitive test of $f_W/\Lambda$ and/or $f_{WW}/\Lambda$, but it is
not sensitive to $f_B/\Lambda^2$ and $f_{BB}/\Lambda^2$ \cite{ZKHY03}.
The present process $\gamma\gamma\to ZZ$ thus provides complementary
information about $f_{BB}/\Lambda^2$.
We note that the signal process is via an $s$-channel scalar exchange, so that
only the two photons with same helicities contribute,
while the leading SM backgrounds come from one-loop box diagrams
and thus are of all partial wave contributions \cite{Jikia}.
In our background calculations, we take $m_t=174$ GeV.

\subsection{The case of polarized photon colliders}

\begin{figure}[tb]
\includegraphics[width=12truecm,clip=true]{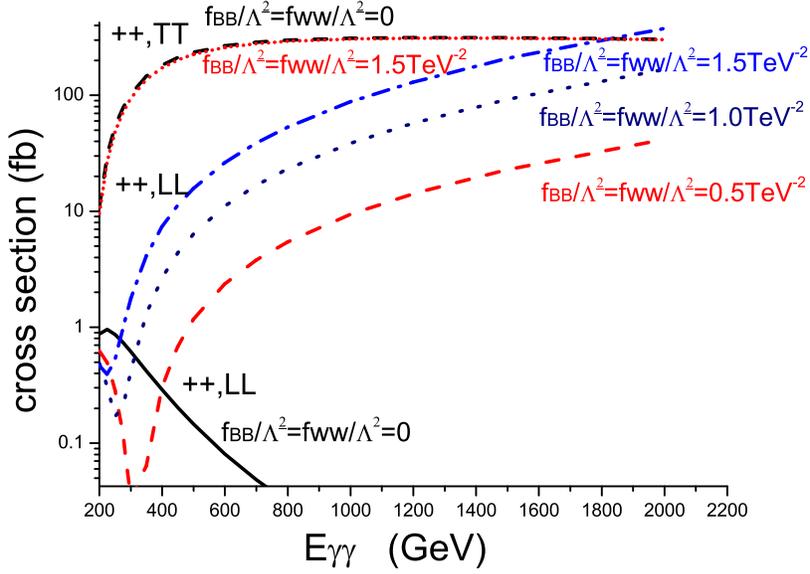}
\null\vspace{-0.5cm}
\caption{Energy dependence of the signal and background cross sections for
$\gamma_+\gamma_+\to Z_LZ_L$ and $\gamma_+\gamma_+\to Z_TZ_T$ in the case of $m_H=115$ GeV.
The symbol $+$ labels the photon
helicity, and the symbols $++,LL$ and $++,TT$ stand for $\gamma_+\gamma_+\to Z_LZ_L$ and
$\gamma_+\gamma_+\to Z_TZ_T$, respectively. For simplicity, we take
$f_{BB}/\Lambda^2=f_{WW}/\Lambda^2\equiv f/\Lambda^2$. The curves with $f/\Lambda^2=0$ are the
SM backgrounds.}
\label{++TTLL}
\end{figure}

As noted  above, the like-sign  helicities for the two photon beams are
preferred by the signal. We thus first consider the case of polarized photon colliders.
It is  instructive to examine the final state $Z$ bosons with
specific  polarizations and see the comparison for the signal and SM
background expectation.
We  illustrate this by plotting the energy dependence of the cross sections with
representative values of the anomalous couplings for various $Z$
polarizations in FIG.~\ref{++TTLL}.
We label the photon circular polarizations by $\pm\pm$,
and the longitudinal (transverse) polarizations of the
final state $Z$ bosons by $LL\ (TT)$.
The $\gamma_+\gamma_+\to Z_LZ_L$ cross section $\sigma(++LL)$ and the
$\gamma_+\gamma_+\to Z_TZ_T$ cross sections $\sigma(++TT)$ are shown
as functions of the collision energy $E_{\gamma\gamma}$ for $m_H=115$ GeV.
The  cross section for $Z_L Z_T$ is so small that we have  ignored it.

We first look at the SM cross sections for longitudinally and
transversely polarized $ZZ$ labeled by
$\sigma(++TT,f/\Lambda^2=0)$ and $\sigma(++LL,f/\Lambda^2=0)$, respectively.
At high energies, $\sigma(++TT,f/\Lambda^2=0)$ is dominantly from the $W^\pm$-loop
diagrams \cite{Jikia} in which the vertices emitting
the $Z_T$ boson are momentum-dependent. So $\sigma(++TT,f/\Lambda=0)$ is large,
and it does not fall off with $E_{\gamma\gamma}$ in the energy range shown in FIG.~\ref{++TTLL}.
The situation of $\sigma(++LL,f/\Lambda^2=0)$ is different. It is easier to understand it by
invoking the equivalence theorem for Goldstone bosons and longitudinal weak bosons \cite{ET}.
According to the equivalence theorem, $Z_L$ can be treated as a would-be Goldstone boson
$z^0$. The vertices emitting $z^0$ in the $W^\pm$-loop diagrams and the $H z^0 z^0$
vertex in the diagram with an $s$-channel Higgs boson are all momentum independent, and the
$H z^0 z^0$ coupling strength of $\lambda\propto m_H^2/v^2$ ($v$=246 GeV)
is weak for a light Higgs with $m_H$ smaller than $v$.
So $\sigma(++LL,f/\Lambda^2=0)$ is
significantly smaller than $\sigma(++TT,f/\Lambda^2=0)$ and it falls off rapidly with
$E_{\gamma\gamma}$ as is seen in the figure.
This implies  that the SM background is mainly $\sigma(++TT,f/\Lambda^2=0)$,
especially in the energy region above 300 GeV.

Next we look at the cross sections with $f/\Lambda^2\ne 0$. From FIG.~\ref{++TTLL} we see that
$\sigma(++TT, f/\Lambda^2\ne 0)$ is very close to the SM cross section
$\sigma(++TT,f/\Lambda^2=0)$. This means that the signal cross section
$\sigma_S(++TT)\equiv \sigma(++TT, f/\Lambda^2\ne 0)-\sigma(++TT,f/\Lambda^2=0)$ defined in
Eq.~(\ref{sigmaBsigmaS}) is negligibly small. The signal of $f/\Lambda^2\ne 0$ is thus
dominated by the $++LL$ channel. Due to the momentum
dependence of the longitudinal polarization vector, the anomalous coupling $f/\Lambda^2\ne 0$ causes
extra energy-dependence of $\sigma(++LL, f/\Lambda^2\ne 0)$ \cite{ZKHY03}. We see that
$\sigma(++LL, f/\Lambda^2\ne 0)$ increases rapidly with $E_{\gamma\gamma}$ (below the new physics
scale  $\Lambda$), and it becomes larger than the SM cross section  $\sigma(++LL, f/\Lambda^2=0)$
at high energies. Therefore the signal cross section
$\sigma_S(++LL)\equiv \sigma(++LL, f/\Lambda^2\ne 0)- \sigma(++LL, f/\Lambda^2=0)$
is large and increases very rapidly with $E_{\gamma\gamma}$. We note from FIG.~\ref{++TTLL}
that there is destructive interference between the SM amplitude and that of
$(++LL, f/\Lambda^2\ne 0)$. The interference would become constructive if we flip the
sign of $f_n$. However, the effects are only at lower energies
$E_{\gamma\gamma}\le 300$ GeV and are nevertheless rather small.
The interference effects at high energies are essentially diminished.
As seen in FIG.~\ref{++TTLL}, the SM background of $Z_LZ_L$ production falls off very rapidly
at higher energies and the signal rate increases  with respect to the anomalous couplings
as $(f/\Lambda^2)^2$. One may wonder if we should include the dimension-8 operators as well
which is of $O(1/\Lambda^4)$ and is important at high energies.
In fact, our approximation is well justified. For the dimension-8 operators,
the linear terms proportional to  $1/\Lambda^4$
arise from the interference with the SM amplitude, which are much smaller for the reason stated above.

At higher energies, the signal
cross section $\sigma_S(++LL, f/\Lambda^2\ne 0)$ may exceed the SM background
$\sigma(++TT,f/\Lambda^2=0)$. However, in the energy region below 1.5 TeV, the signal
$\sigma_S(++LL,f/\Lambda^2\ne 0)$ is still smaller than the background $\sigma(++TT,f/\Lambda^2=0)$.
In addition to $\sigma(++TT,f/\Lambda^2=0)$ considered above, there is also
SM backgrounds $\sigma(\pm\mp TT,f/\Lambda^2=0)$ from $\gamma_\pm \gamma_\mp$ collisions, which are
comparable to the $\sigma(++TT,f/\Lambda^2=0)$ \cite{Jikia}.
Over all, we see that  the SM background is substantially larger than
the signal except at very high energies.
Thus we need to develop suitable kinematic cuts to suppress
the SM backgrounds, as we discuss next.

\begin{figure}[t]
\includegraphics[width=17truecm,clip=true]{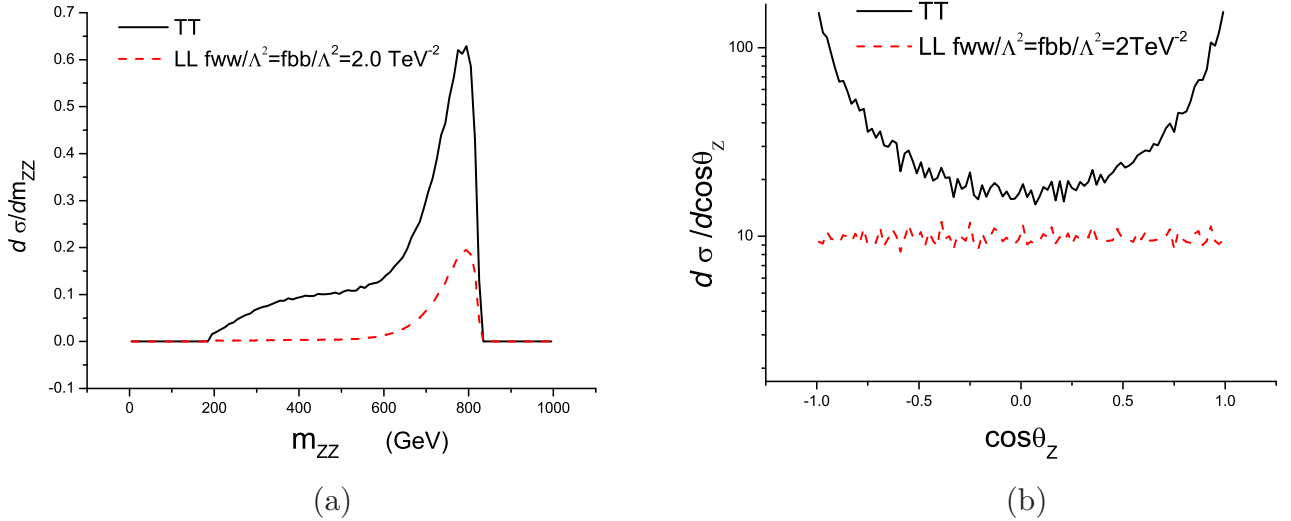}
\null\vspace{-0.2cm}
\caption{Differential distributions for (a) the invariant mass $M_{ZZ}$ and
(b) $\theta_Z$ at $\sqrt{s_{ee}}=1$ TeV in the $e^+e^-$ collision system after imposing the
cut (\ref{MZZ}). The solid curve is the SM background $\sigma(TT,f/\Lambda^2=0)$
and the dashed curve is the signal $\sigma(++LL,f/\Lambda^2=2.0~{\rm TeV}^{-2})$.}
\label{M(ZZ)}
\end{figure}

Our first cut is for suppressing the $\gamma_+\gamma_-$ contribution.
We know that the colliding photon energy will peak around $80\%$ of the $e^+e^-$ colliding energy
with a tail in the low energy region  if we take the polarization satisfying
$2\lambda_eP_c=-1$ [cf. FIG.~1(a)] \cite{Ginzburg}.
The corresponding $\omega/E_e$ distribution of the mean
helicity $\lambda_r$ of the colliding photon for various polarizations
has been given in Ref.~\cite{TESLA6} which is shown in Fig.~1(b).
We see from curve $a$ in Fig.~1(b) ($2\lambda_eP_c=-1$) that the mean helicity is nearly
$\lambda_r\approx +1$ in the region $0.65<\omega/E_e<0.8$. Therefore we can envision
to impose this condition to strongly suppressed the background.
In FIG.~\ref{M(ZZ)}(a), we plot the $M_{ZZ}$ distribution for the signal
$\sigma(++LL,f/\Lambda^2=2~{\rm TeV}^{-2})$  and the background
$\sigma(TT,f/\Lambda^2=0)$ at $\sqrt{s_{ee}}=1$ TeV. This motivates us to
introduce a cut on the invariant mass of the final state $ZZ$
\begin{eqnarray}                               
M_{ZZ}>0.65\sqrt{s_{ee}}.
\label{MZZ}
\end{eqnarray}
\null\noindent
Numerically, we find that the cut (\ref{MZZ}) reduces the background
$\sigma(TT,f/\Lambda^2=0)$ from 115 fb to 72 fb
down by $37\%$, but the signal $\sigma(++LL,f/\Lambda^2=2~{\rm TeV}^{-2})$ only
from 22 fb to 20 fb. After imposing the cut in
(\ref{MZZ}), the SM background contains mainly
$\sigma(++TT,f/\Lambda^2=0)$ shown in FIG.~\ref{++TTLL}.

Our second cut concerns the angular distribution of the final state $Z$ bosons.
In FIG.~\ref{M(ZZ)}(b), we plot the $\theta_Z$ distributions of the signal
$\sigma(++LL,f/\Lambda^2=2~{\rm TeV}^{-2})$  and
the background $\sigma(++TT,f/\Lambda^2=0)$ at a $\sqrt{s_{ee}}=1$ TeV linear collider,
where $\theta_Z$ is the polar angle of the final state $Z$ boson with respect to
the colliding beam $e^-$ axis in the center of mass frame of $ZZ$.  As expected,
$d\sigma(LL,f/\Lambda^2=2~{\rm TeV}^{-2})/d\cos\theta_Z$ is isotropic,
while $d\sigma(TT,f/\Lambda^2=0)/d\cos\theta_Z$ is strongly forward and backward.
Therefore we impose
\begin{eqnarray}                             
-0.5<\cos\theta_Z<0.5.
\label{theta_Z}
\end{eqnarray}
We find that, after this cut, $\sigma(TT,f/\Lambda^2=0)$ is reduced
from 72 fb to 16 fb, while $\sigma(LL,f/\Lambda^2=2~{\rm TeV}^{-2})$
changes from 20 fb to 9 fb.

It is known that the
angular distribution for a fermion from $Z$ decay goes like
\begin{eqnarray}
\nonumber
{d\sigma\over d\cos\theta_f} \propto \left\{
\begin{array}{ll}
1 + \cos^2\theta_f \qquad   {\rm  for}\ Z_T, & \\
\sin^2\theta_f \qquad   ~~~~~~{\rm  for}\ Z_L, &
\end{array}
\right.
\end{eqnarray}
where $\theta_f$ is the polar angle for a final state fermion $f$ in the rest frame of $Z$
with respect to the $Z$ momentum direction in the center of mass frame of $ZZ$.
We thus propose a third cut
\begin{eqnarray}          
|\cos\theta_f | < 0.5.
\label{DeltaEcut}
\end{eqnarray}
In the example of $\sqrt{s_{ee}}=1$ TeV, this cut reduces the background cross section
from 16 fb to 3.5 fb, while it reduces
$\sigma(LL,f/\Lambda^2=2~{\rm TeV}^{-2})$ from 9 fb to 5 fb.
To measure the effectiveness of our cuts, we define a double-ratio
\begin{eqnarray}                 
\displaystyle{r\equiv \frac{\sigma_{LL}({\rm after~cut}))/
\sigma_{LL}({\rm before~cut})}
{\sigma_{TT}({\rm after~cut}))/\sigma_{TT}({\rm before~cut})}}.
\label{effi}
\end{eqnarray}
The larger the value of this ratio is, the more effective the cuts are in terms
of the signal enhancement over the background suppression,
The effects of the three cuts are summarized in TABLE \ref{cuts}
for $\sqrt{s_{ee}}=1$ and 3 TeV.

\begin{table}[h]
\caption{The cut effectiveness $r$ as defined in Eq.~(\ref{effi}), in terms of the
effects of the cuts (\ref{MZZ}), (\ref{theta_Z}), and (\ref{DeltaEcut}) in the case of the
$\sqrt{s_{ee}}=1$ TeV ILC with $f/\Lambda^2=2~{\rm TeV}^{-2}$ and
the case of the $\sqrt{s_{ee}}=3$ TeV CLIC with $f/\Lambda^2=1~{\rm TeV}^{-2}$. The symbols
$\sigma_{LL}$ and $\sigma_{TT}$ stand for $\sigma(++LL,f/\Lambda^2\ne 0)$ and
the background cross section, respectively.
}
\tabcolsep 8pt
\null\vspace{0.2cm}
\begin{tabular}{cccccc}
\hline\hline
&&no cuts &cut (\ref{MZZ})&cuts (\ref{MZZ})+(\ref{theta_Z})
&cuts (\ref{MZZ})+(\ref{theta_Z})+(\ref{DeltaEcut})\\
\hline\\
$\sqrt{s_{ee}}=1$ TeV&$\sigma_{LL}$ (fb)&22&20&9.2& 4.9\\
&$\sigma_{TT}$ (fb)&115&72&16& 3.5\\
&$r$&$\--$&${90\%}/{63\%}\approx1.4$&${47\%}/{22\%}\approx 2.1$&${54\%}/{22\%}\approx 2.5$\\
&&&&\\
$\sqrt{s_{ee}}=3$ TeV&$\sigma_{LL}$ (fb)&58&51&24&12.2\\
&$\sigma_{TT}$ (fb)&190&85&6.5&1.1\\
&$r$&$\--$&$88\%/45\%\approx 2.0$&$47\%/7.6\%\approx 6.2$&$51\%/17\%\approx 3.0$\\
\hline\hline
\label{cuts}
\end{tabular}
\end{table}

Next we consider the actual detection of the final state $Z$ bosons  via the
decay products. We only consider the fully reconstructable modes $Z\to \ell^+\ell^-\
(\ell=e,\ \mu)$  and $ jj$, with branching ratios  \cite{PDG}
\begin{eqnarray*}
B(Z\to e^+e^-)=B(Z\to\mu^+\mu^-)\approx 3.36\% ,\ \  B(Z\to jj)\approx 70\% .
\end{eqnarray*}
For the two final state $Z$ bosons, we can have the pure leptonic modes $ZZ\to \ell^+\ell^-\
\ell^+\ell^-$,  semi-leptonic modes $ZZ\to \ell^+\ell^-\ jj$, and the hadronic modes $ZZ\to jj\ jj$.
We require the reconstruction of $Z$ bosons by the invariant mass of the decay products
$M_Z\approx M(\ell^+\ell^-),\ M(jj)$. For the hadronic mode,
some care needs to be taken due to the potentially large background
$\gamma\gamma\to W^+W^-\to jjjj$. It has been emphasized that it is important for
the detector to be able to distinguish the hadronic decays of the $Z$ and $W$
from their mass reconstruction \cite{WZ}.
As a conservative estimate for the reconstruction efficiency,
we include another factor $50\%$ for the $ZZ$ hadronic modes.
Then the total detection efficiency for the final state $ZZ$ under consideration is
\begin{eqnarray}
\epsilon=(3.36\%+3.36\%)^2+2\times(3.36\%+3.36\%)\times 70\%+
(70\%)^2\times 50\%\approx 34\%.
\label{epsilon}
\end{eqnarray}

\begin{figure}[tb]
\includegraphics[width=17truecm,clip=true]{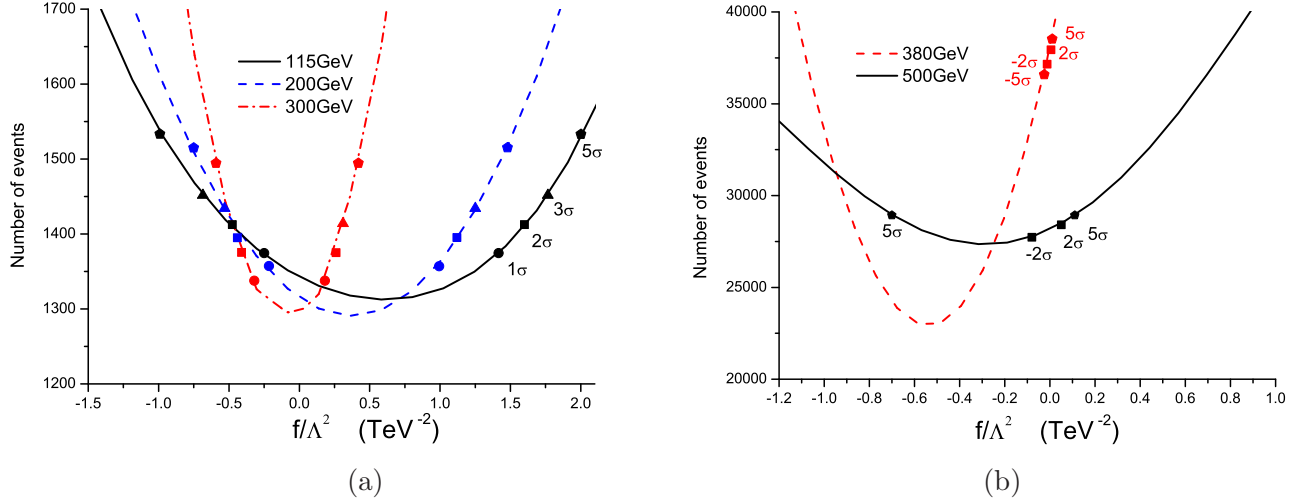}
\null\vspace{-0.4cm}
\caption{Numbers of events of $\gamma\gamma\to ZZ~(Z\to e^+e^-,\mu^+\mu^-,jj)$
versus $f/\Lambda^2$: (a) for $m_H=115$ GeV,~200 GeV and 300 GeV
at a $\sqrt{s_{ee}}=500$ GeV polarized ILC with the cuts (\ref{MZZ}), (\ref{theta_Z}) and
(\ref{DeltaEcut}) imposed, (b) for $m_H=300$ GeV at a $\sqrt{s_{ee}}=500$ GeV polarized ILC
(solid curve) and a $\sqrt{s_{ee}}=380$ GeV polarized ILC (dashed curve) without imposing the cuts
(\ref{MZZ}), (\ref{theta_Z}) and (\ref{DeltaEcut}).
The values of $f/\Lambda^2$ corresponding to $1\sigma$, $2\sigma$, $3\sigma$ and $5\sigma$ statistical deviations [cf.~Eq.~(\ref{N})] are shown on each curve by the
$f/\Lambda^2$-coordinates of the bullet, square, triangle and asterisk, respectively.}
\label{AAZZ500P}
\end{figure}

For comparison with the LHC results of Ref.~\cite{ZKHY03},
we make the same assumption
$f_{BB}/\Lambda^2=f_{WW}/\Lambda^2 \equiv f/\Lambda^2$
for illustration.
The event rates can be calculated using Eqs.~(\ref{dN'}) and (\ref{epsilon}).
Assuming an integrated luminosity of  1 ab$^{-1}$,
we obtain the number of events  in FIG.~\ref{AAZZ500P}(a)  for
$m_H=115,\ 200$ and $300~{\rm GeV}$ at a $\sqrt{s_{ee}}=500$ GeV polarized ILC
with the cuts (\ref{MZZ}), (\ref{theta_Z}) and (\ref{DeltaEcut}).
The symbols of the bullet, square, triangle and asterisk on each curve
mark the values corresponding to $1\sigma$, $2\sigma$, $3\sigma$ and $5\sigma$
statistical significance  $\sigma_{stat}$ of Eq.~(\ref{N}), respectively.
The numbers of events are above
1300 for $m_H=115\-- 300$ GeV.  It is not so sensitive to $m_H$, but
varies quite sensitively with respect to the  value of $f/\Lambda^2$.

At this kind of energy, in the case of $m_H=115-200$ GeV, the signal cross section
is comparable to that of the SM background after the cuts.
There is considerable interference between the two amplitudes, which causes the numbers
of events asymmetric near $f/\Lambda^2\approx 0$.
If  no $f/\Lambda^2\ne 0$ signal effect is observed at the ILC, it
would lead to sensitive bounds on the anomalous couplings.
From FIG.~\ref{AAZZ500P}(a)  for  $m_H=115-200$ GeV,
we obtain the $2\sigma$ and $3\sigma$ bounds for
$f/\Lambda^2$ and $g_{H\gamma\gamma}$ [cf.~Eq.~(\ref{g})]:
\begin{eqnarray}                          
&&\sqrt{s_{ee}}=500~{\rm GeV~polarized~ILC},\ m_H=115\--200~{\rm GeV}:
\hspace{5cm}\nonumber\\
&&~~2\sigma:~~-0.48~{\rm TeV}^{-2}<f/\Lambda^2<1.6~{\rm TeV}^{-2},
~~-0.019~{\rm TeV}^{-1}<g_{H\gamma\gamma}<0.0058~{\rm TeV}^{-1},\nonumber\\
&&~~3\sigma:~~-0.68~{\rm TeV}^{-2}<f/\Lambda^2<1.8~{\rm TeV}^{-2},
~~-0.022~{\rm TeV}^{-1}<g_{H\gamma\gamma}<0.0082~{\rm TeV}^{-1}.
\label{500Pfg}
\end{eqnarray}
This $2\sigma$ constraint on $g_{H\gamma\gamma}$
is comparable to  those in the literature \cite{HZZ} from the other
processes. From the  results in Ref.~\cite{ZKHY03}
we can see that the $2\sigma$ ($3\sigma$) constraints
on $f_{WW}/\Lambda^2$ and  $g_{H\gamma\gamma}$
obtained from $W^+W^+$ scattering at the LHC are
\begin{eqnarray}
|f_{WW}/\Lambda^2|\le 2.2\ {\tev^{-2}}\ \ (3.0\ {\tev^{-2}}), \quad
|g_{H\gamma\gamma}|<0.027~{\rm TeV}^{-1}\ \ (0.036~{\rm TeV}^{-1}).
\end{eqnarray}
Comparing with our results in Eq.~(\ref{500Pfg}) for $\gamma\gamma\to ZZ$,
 we see that   the upper bound would be improved over the LHC results
by a factor about 1.5, and the lower bound  by roughly a factor of 4.

The case of $m_H=300$ GeV at the 500 GeV ILC is rather special.
Even with the cut (\ref{MZZ}), 
the tail of the resonance still enhances the contribution in the %
$s$-channel Higgs diagrams (e.g. FIG.~\ref{phph-ZZ}).
On the other hand, the SM background via the $s$-channel Higgs boson
exchange does not increase as much. Consequently,
the number of events for $f/\Lambda^2\ne 0$ is significantly larger for
$m_H=300$ GeV than those for $m_H=115-200$ GeV.
Thus the test for $m_H=300$ GeV is much more sensitive
than that for $m_H=115-200$ GeV.
Because the resonance enhancement is more significant in the signal amplitude of
FIG.~\ref{phph-ZZ} than in the SM amplitude, we may even consider relaxing
the cuts (\ref{MZZ}), (\ref{theta_Z}), and (\ref{DeltaEcut}).
Although the SM background is less suppressed without imposing the cuts,
the signal to background ratio can still be improved. 
For illustration, the result for $m_H=300$ GeV without those
cuts is plotted as the solid curve in FIG.~\ref{AAZZ500P}(b). When including
the resonant signal, the number of events can be increased by an order of
magnitude.
The  sensitivity is much improved.
The $5\sigma$ sensitivity can constrain the couplings in the range
\begin{eqnarray}                         
&&\sqrt{s_{ee}}=500~{\rm GeV~polarized~ILC~(no~cuts)},~m_H=300~{\rm GeV}:
\hspace{5cm}\nonumber\\
&&~~5\sigma:~~ -0.7~{\rm TeV}^{-2}<f/\Lambda^2<0.11~{\rm TeV}^{-2},~~
-0.0013~{\rm TeV}^{-1}<g_{H\gamma\gamma}<0.0085~{\rm TeV}^{-1}.
\label{500Pfg'}
\end{eqnarray}

In practice, if a light Higgs boson with certain $m_H$ is found at the LHC,
we may consider tuning the colliding photon energy $E_{\gamma\gamma}$  to be
at the resonance value of the mass since the energy in the first phase of the ILC
can be tuned between 200 GeV and 500 GeV. Then the event rate for the
$f/\Lambda^2\ne 0$ amplitude shown in FIG.~\ref{phph-ZZ} is maximally enhanced,
and we can reach the optimal sensitivity. This works for the cases with $m_H>2M_Z$.
Take the case of $m_H=300$ GeV as an example again.
One can tune the ILC energy to be $\sqrt{s_{ee}}=380$ GeV so that the peak of
$E_{\gamma\gamma}$ is at 300 GeV.
The calculated result in this case is plotted as the dashed curve in FIG.~\ref{AAZZ500P}(b).
The $5\sigma$ sensitivity range is
\begin{eqnarray}                         
&&\sqrt{s_{ee}}=380~{\rm GeV~polarized~ILC~(without~cuts)},~m_H=300~{\rm GeV}:
\hspace{5cm}\nonumber\\
&&~~5\sigma:~ -0.025~{\rm TeV}^{-2}<f/\Lambda^2<0.010~{\rm TeV}^{-2},~
-0.00012~{\rm TeV}^{-1}<g_{H\gamma\gamma}<0.00030~{\rm TeV}^{-1}.
\label{380Pfg}
\end{eqnarray}
We see that the sensitivity can be very high. To put it into perspective, if $f$ is naturally
the order of unity, then the physical scale probed can be as high as $\Lambda \sim 10\ {\tev}$.
Note that we have only taken into account the statistical error here. With the systematic error
(depending on the property of the detectors), the actual sensitivity may be accordingly lower.

\begin{figure}[tbh]
\includegraphics[width=17truecm,clip=true]{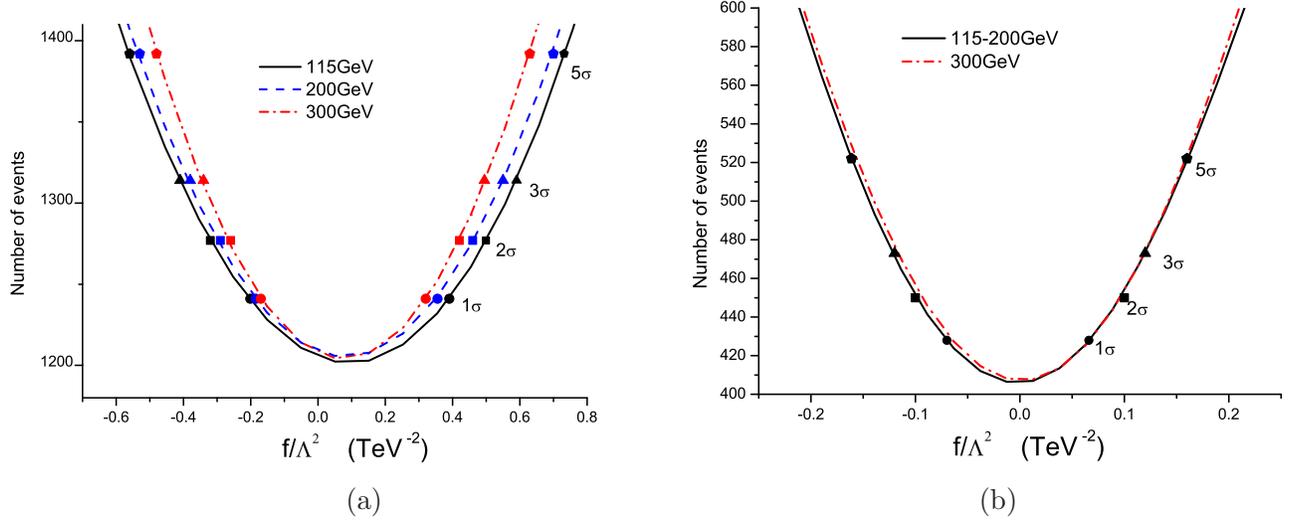}
\null\vspace{-0.4cm}
\caption{Numbers of events of $\gamma\gamma\to ZZ~(Z\to e^+e^-,\mu^+\mu^-,jj)$
versus $f/\Lambda^2$ for $m_H=115$ GeV, 200 GeV and 300 GeV
 with the cuts (\ref{MZZ}), (\ref{theta_Z}) and
(\ref{DeltaEcut}): (a) at a $\sqrt{s_{ee}}=1$ TeV polarized ILC, (b) at a $\sqrt{s_{ee}}=3$ TeV
polarized CLIC.
The values of $f/\Lambda^2$ corresponding to $1\sigma$, $2\sigma$, $3\sigma$ and $5\sigma$ statistical
deviations [cf.~Eq.~(\ref{N})] are shown on each curve by the
$f/\Lambda^2$-coordinates of the bullet, square, triangle and asterisk, respectively.}
\label{AAZZ1-3P}
\end{figure}

Next we consider the cases of $\sqrt{s_{ee}}=1$ TeV polarized ILC and $\sqrt{s_{ee}}=3$ TeV
polarized CLIC for $m_H=115\--300$ GeV. The obtained results are plotted in FIG.~\ref{AAZZ1-3P}.
We see that the SM backgrounds are almost independent of $m_H$ as the values shown
at $f/\Lambda^2 = 0$. Furthermore, for $\sqrt{s_{ee}}=3$ TeV in FIG.~\ref{AAZZ1-3P}(b),
 the event rates are essentially symmetric near $f/\Lambda^2=0$, due to the large signal
 rate and small interference with the SM amplitude. In the absence of a signal observation,
FIG.~\ref{AAZZ1-3P}(a) and (b) lead to the following $2\sigma$ and $3\sigma$
statistical  bounds
\begin{eqnarray}                          
&&\sqrt{s_{ee}}=1~{\rm TeV~polarized~ILC~(with~cuts)},~m_H=115\--300~{\rm GeV}:
\hspace{5cm}\nonumber\\
&&~~2\sigma:~~-0.32~{\rm TeV}^{-2}<f/\Lambda^2<0.50~{\rm TeV}^{-2},~~
-0.0061~{\rm TeV}^{-1}<g_{H\gamma\gamma}<0.0048~{\rm TeV}^{-1},\nonumber\\
&&~~3\sigma:~~-0.40~{\rm TeV}^{-2}<f/\Lambda^2<0.60~{\rm TeV}^{-2},~~
-0.0073~{\rm TeV}^{-1}<g_{H\gamma\gamma}<0.0054~{\rm TeV}^{-1}.
\label{1000Pfg}
\end{eqnarray}

\null\vspace{-1.8cm}
\begin{eqnarray}                           
&&\sqrt{s_{ee}}=3~{\rm TeV~polarized~CLIC~(with~cuts)},~m_H=115\--300~{\rm GeV}:
\hspace{5cm}\nonumber\\
&&~~2\sigma:~~-0.10~{\rm TeV}^{-2}<f/\Lambda^2<0.10~{\rm TeV}^{-2},~~
-0.0012~{\rm TeV}^{-1}<g_{H\gamma\gamma}<0.0012~{\rm TeV}^{-1},\nonumber\\
&&~~3\sigma:~~-0.12~{\rm TeV}^{-2}<f/\Lambda^2<0.12~{\rm TeV}^{-2},~~
-0.0015~{\rm TeV}^{-1}<g_{H\gamma\gamma}<0.0015~{\rm TeV}^{-1}.
\label{3000Pfg}
\end{eqnarray}
We see that the present tests lead to very high sensitivities. The constraints on
$g_{H\gamma\gamma}$ are all of the order of $10^{-3}$ TeV$^{-1}$.
Compared with the $3\sigma$ constraints $|f_{WW}/\Lambda^2|\le 2.9$ TeV$^{-2}$ and
$|g_{H\gamma\gamma}|<0.036~{\rm TeV}^{-1}$ obtained from the results of $W^+W^+$
scattering at the LHC in Ref.~\cite{ZKHY03},
the present sensitivities at $\sqrt{s_{ee}}=1$ TeV and
$\sqrt{s_{ee}}=3$ TeV are improved by roughly a factor of 5 and 24, respectively.

\subsection{The case of unpolarized photon colliders}

\begin{figure}[tbh]
\includegraphics[width=15truecm,clip=true]{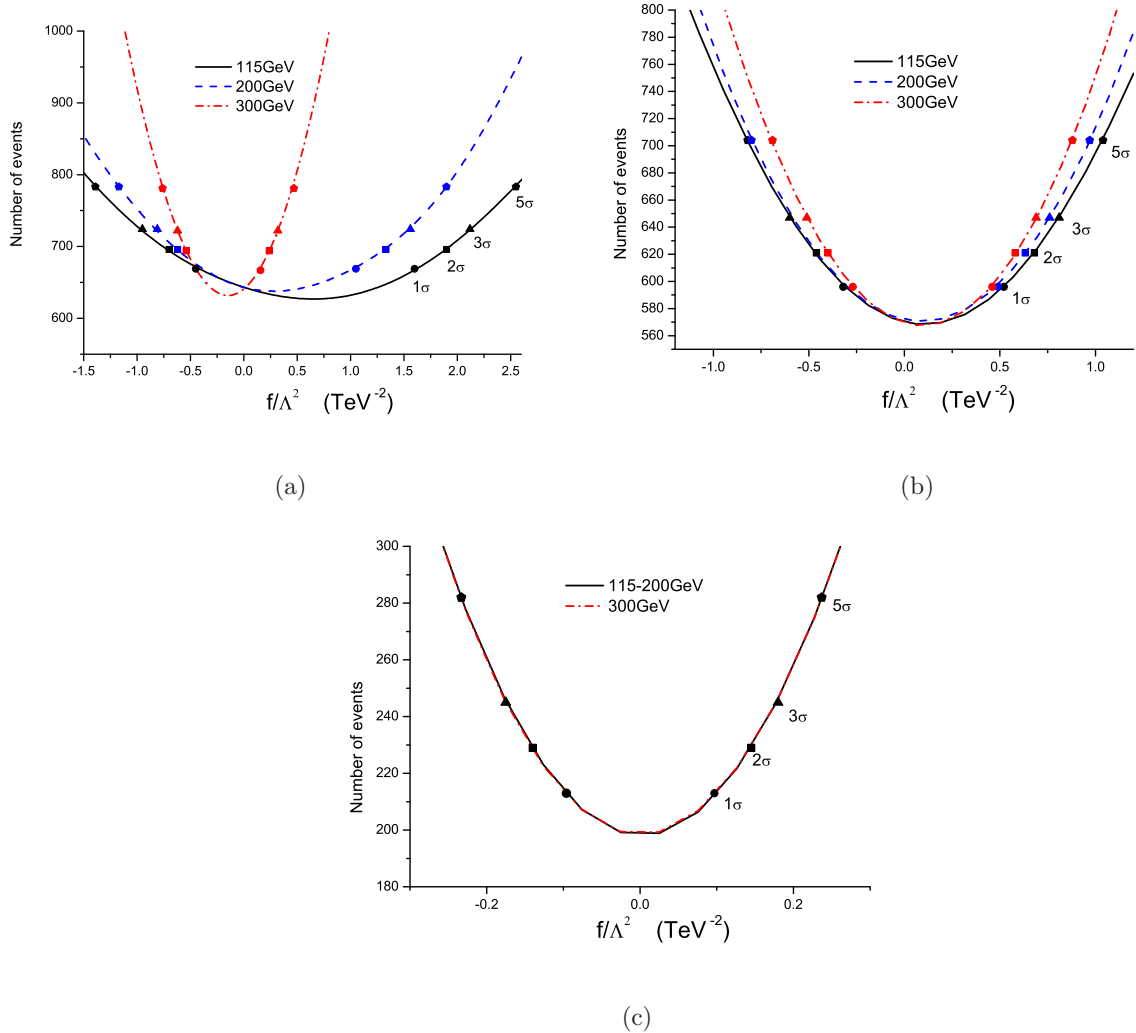}
\null\vspace{-0.3cm}
\caption{Numbers of events of $\gamma\gamma\to ZZ~(Z\to e^+e^-,\mu^+\mu^-,jj)$
versus $f/\Lambda^2$ for $m_H=115$ GeV, 200 GeV and 300 GeV
 with the cuts (\ref{MZZ}), (\ref{theta_Z}) and
(\ref{DeltaEcut}) at unpolarized linear collders: (a) at a $\sqrt{s_{ee}}=500$ GeV ILC,
(b) at a $\sqrt{s_{ee}}=1$ TeV ILC, and (c) at a $\sqrt{s_{ee}}=3$ TeV CLIC.
The values of $f/\Lambda^2$ corresponding to $1\sigma$, $2\sigma$, $3\sigma$ and $5\sigma$ statistical
deviations [cf.~Eq.~(\ref{N})] are shown on each curve by the
$f/\Lambda^2$-coordinates of the bullet, square, triangle and asterisk, respectively.}
\label{AAZZU}
\end{figure}

The production cross section at the unpolarized photon colliders
can be calculated with Eq.~(\ref{sigma-unp}).
In the unpolarized case, the colliding photon energy distribution
$E_{\gamma\gamma}$ is less sharply peaked as seen in FIG.~{\ref{mean}}(a).
This reduces the  sensitivity with respect to the polarized case.
The calculated results for $m_H=115,~130,~200,~300$ GeV
at the $\sqrt{s_{ee}}=500$ GeV ILC, the 1 TeV ILC, and the 3 TeV CLIC with the
same cuts (\ref{MZZ}), (\ref{theta_Z}) and
(\ref{DeltaEcut}) are plotted in FIG.~\ref{AAZZU}.

From FIG.~\ref{AAZZU}(a) we see that the $m_H=300$ GeV case is again especially sensitive
due to the nearly on-shell resonance effect as discussed earlier.
For a lighter mass of $m_H=115$ and 200 GeV, the $2\sigma$ and $3\sigma$
 sensitivity bounds at the 500 GeV unpolarized ILC are
\begin{eqnarray}                          
&&\sqrt{s_{ee}}=500~{\rm GeV~unploarized~ILC~(with~cuts)},~m_H=115\--200~{\rm GeV}:
\hspace{5cm}\nonumber\\
&&~~2\sigma:~~-0.7~{\rm TeV}^{-2}<f/\Lambda^2<1.9~{\rm TeV}^{-2},~~
-0.023~{\rm TeV}^{-1}<g_{H\gamma\gamma}<0.0085~{\rm TeV}^{-1},\nonumber\\
&&~~3\sigma:~~-1.0~{\rm TeV}^{-2}<f/\Lambda^2<2.1~{\rm TeV}^{-2},~~
-0.025~{\rm TeV}^{-2}<g_{H\gamma\gamma}<0.012~{\rm TeV}^{-2}.
\label{500Ufg}
\end{eqnarray}
These sensitivities are lower than those in Eq.~(\ref{500Pfg}) in the polarized case
by roughly a factor of 1.2.

From FIG.~\ref{AAZZU}(b) we see that the $2\sigma$ and $3\sigma$ testing sensitivities
at the 1 TeV unpolarized ILC are
\begin{eqnarray}                          
&&\sqrt{s_{ee}}=1~{\rm TeV~unpolarized~ILC~(with~cuts)},~m_H=115\--300~{\rm GeV}:
\hspace{4.5cm}\nonumber\\
&&~~2\sigma:~~-0.46~{\rm TeV}^{-2}<f/\Lambda^2<0.68~{\rm TeV}^{-2},~~
-0.0082~{\rm TeV}^{-1}<g_{H\gamma\gamma}<0.0056~{\rm TeV}^{-1},\nonumber\\
&&~~3\sigma:~~-0.60~{\rm TeV}^{-2}<f/\Lambda^2<0.81~{\rm TeV}^{-2},~~
-0.0098~{\rm TeV}^{-2}<g_{H\gamma\gamma}<0.0073~{\rm TeV}^{-2}.
\label{1000Ufg}
\end{eqnarray}
These are lower than that in the polarized case [cf.~Eq.~(\ref{1000Pfg})] by
roughly a factor of 1.3.

From FIG.~\ref{AAZZU}(c) we see that the $2\sigma$ and $3\sigma$ testing sensitivity at the 3 TeV
unpolarized CLIC are
\begin{eqnarray}                          
&&\sqrt{s_{ee}}=3~{\rm TeV~unploarized~CLIC~(with~cuts)},~m_H=115\--300~{\rm GeV}:
\hspace{4.5cm}\nonumber\\
&&~~2\sigma:~~-0.14~{\rm TeV}^{-2}<f/\Lambda^2<0.14~{\rm TeV}^{-2},~~
-0.0017~{\rm TeV}^{-1}<g_{H\gamma\gamma}<0.0017~{\rm TeV}^{-1},\nonumber\\
&&~~3\sigma:~~-0.18~{\rm TeV}^{-2}<f/\Lambda^2<0.18~{\rm TeV}^{-2},~~
-0.0022~{\rm TeV}^{-1}<g_{H\gamma\gamma}<0.0022~{\rm TeV}^{-1}.
\label{3000Ufg}
\end{eqnarray}
These are lower than that in the polarized case [cf.~Eq.~(\ref{3000Pfg})] by
roughly a factor of 1.4.

We conclude that suitable polarization of the photon collider does increase the testing sensitivities
relative to the unpolarized case by roughly a factor of $1.2\--1.4$.

\section{Sensitivity to the anomalous couplings from $\bm{\gamma\gamma\to WWWW}$}
\label{wwww}

In order to avoid the large SM background in $\gamma\gamma\to W_TW_T,~Z_TZ_T$,
it was suggested in Refs. \cite{Brodsky,Cheung94L} to make use of the processes
 $\gamma\gamma\to WWWW$ and $\gamma\gamma\to WWZZ$ to
study strongly interacting EWSBM without a light Higgs boson.
The signals and backgrounds in these processes were carefully
studied in Ref.~\cite{Cheung94D}.
Here we explore the anomalous gauge couplings of the
light Higgs boson [cf.~Eqs.~(\ref{eff})$-$(\ref{g})] by the process
\begin{equation}
\gamma\gamma\to WWWW.
\label{AAWWWW}
\end{equation}
It is easy to   show that the $\gamma\gamma\to WWWW$ process is mainly
sensitive to the anomalous coupling $f_{WW}/\Lambda^2$ and $f_W/\Lambda^2$ but not  to
$f_B/\Lambda^2$ and $f_{BB}/\Lambda^2$. This is because that $f_B/\Lambda^2$ and $f_{BB}/\Lambda^2$
appear in $g^{(i)}_{HVV}$'s with coefficients proportional to power(s) of
$\sin\theta_W$ [cf.~Eq.~(\ref{g})].
 In the following, we study  $f_{WW}/\Lambda^2$ and
$f_W/\Lambda^2$ assuming only one dominant at a time.

\begin{figure}[tbh]
\includegraphics[width=7.4truecm,clip=true]{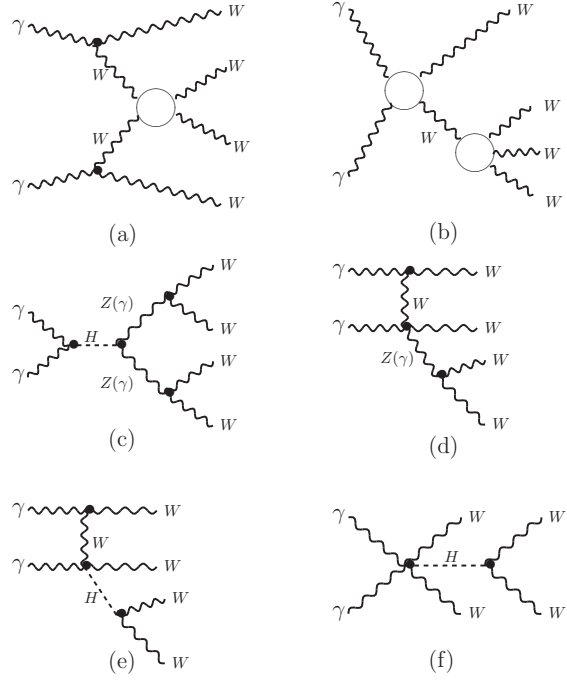}
\caption{Feyman diagrams for the process $\gamma\gamma\to WWWW$.}
\label{AA-WWWW}
\end{figure}

\begin{figure}[tbh]
\includegraphics[width=8truecm,clip=true]{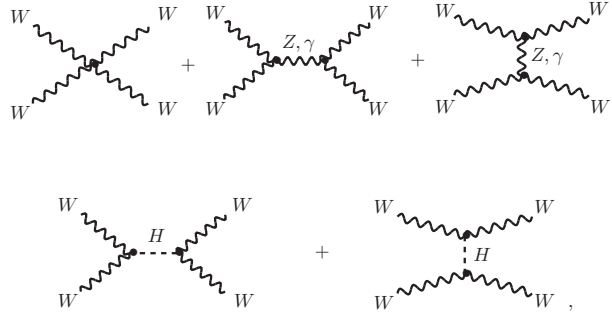}
\caption{Feyman diagrams in the $WWWW$ vertex.}
\label{WWWW}
\end{figure}

\begin{figure}[tbh]
\includegraphics[width=8truecm,clip=true]{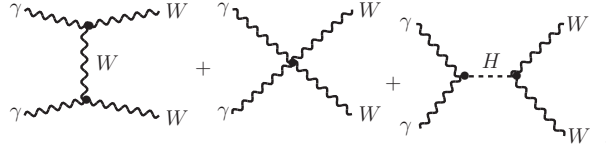}
\caption{Feyman diagrams in the $AAWW$ vertex.}
\label{AAWW}
\end{figure}

The Feynman diagrams for the process of Eq.~(\ref{AAWWWW})
are shown in FIG.~\ref{AA-WWWW}, in which the compact $WWWW$ vertex
collectively stands for the diagrams shown in FIG.~\ref{WWWW},
and the compact $\gamma\gamma WW$ vertex stands for the diagrams shown in FIG.~\ref{AAWW}.
FIG.~\ref{AA-WWWW}(a)$\--$\ref{AA-WWWW}(d) indicate the contributions of the anomalous
couplings $f_{WW}/\Lambda^2$ and $f_W/\Lambda^2$
to $\gamma\gamma\to WWWW$ through its related anomalous $HWW$, $HZZ$, $HZ\gamma$ and $H\gamma\gamma$
interactions [cf.~Eqs.~(\ref{eff})$-$(\ref{g})]. Note that the anomalous coupling $g_{H\gamma\gamma}$
is related to $s^2(f_{BB}+f_{WW})/\Lambda^2$ [cf.~Eq.~(\ref{g})].
Thus for the cases of
$f_{WW}/\Lambda^2$ or $f_W/\Lambda^2$ dominance, FIG.~\ref{AA-WWWW}(c) gives significant
contribution only when $f_{WW}/\Lambda^2$ is not too small.
In the case of $f_W/\Lambda^2$ dominance, i.e.,
$f_{WW}/\Lambda^2,~f_{BB}/\Lambda^2\ll f_W/\Lambda^2$,
the contribution of FIG.~\ref{AA-WWWW}(c) is negligibly small. We note that the cases of
$f_{WW}/\Lambda^2$ dominance or $f_W/\Lambda^2$ dominance
are quite different.
The anomalous coupling $f_{WW}/\Lambda^2$ ($f_W/\Lambda^2$) also gives rise to the more complicated
anomalous $H\gamma WW$ and $H\gamma\gamma WW$ interactions from the nonlinear gauge-field
terms in $\hat{W}_{\mu\nu}$ in ${\cal O}_{WW}$ [cf.~Eq.~(\ref{O})]. These contributions are shown
in FIG.~\ref{AA-WWWW}(e) and \ref{AA-WWWW}(f). Since these terms contain higher powers of the weak
interaction coupling constant $g$, their contributions are much smaller than those from
FIG.~\ref{AA-WWWW}(a)$\--$\ref{AA-WWWW}(d).

For the detection of the four final state
$W$ bosons, we take the hadronic decay  mode $W\to jj$.
Let $\theta$ be the angle between two jets. Experimentally,
the two jets can be resolved
when $\cos\theta<0.8$ \cite{JLC}. At high energy linear
colliders, the $W$ momenta are higher and consequently
the two jets from the decay of a $W$ boson
are collimated along the $W$ moving direction, so that
the condition $\cos\theta<0.8$ can hardly be satisfied.
However, we can still reconstruct the $W$ boson by all the hadronic decay products
in this  wide jet to form a correct mass $M_W$. For
distinguishing two of the final state $W$ bosons, we require the isolation
between any two jets from the $i$-th and $j$-th $W$-boson decays to satisfy
\begin{eqnarray}                       
|\cos\theta_{ij}|<0.8.
\label{theta_ij}
\end{eqnarray}
Here we approximately take the quarks from the $W$ decays as the jets without performing
hadronization and detector  simulation. It turns out that
the requirement (\ref{theta_ij}) slightly  improve the signal to background ratio.
When estimating the detection efficiency, we take again a safety factor  $50\%$ for distinguishing
the final state $W$ bosons from $Z$ bosons. We also take the integrated luminosity of
 1 ab$^{-1}$
to estimate the numbers of events.

We first note that for  the $\sqrt{s_{ee}}=500$ GeV ILC,
the effective colliding photon energy is typically less than 400 GeV, so that
the $\gamma\gamma\to WWWW$ process experiences
severe phase space suppression and  yields only a few events with 1 ab$^{-1}$
luminosity. We will not pursue this case further.
%
%
As for the case of the $\sqrt{s_{ee}}=1$ TeV ILC, with the requirement of Eq.~(\ref{theta_ij}),
our calculation shows that there can be about 200$\--$300 hundred events, and the $2\sigma$
sensitivity for $m_H=115$ GeV is about $-4.5~{\rm TeV}^{-2}<f_W/\Lambda^2<4.5~{\rm TeV}^{-2}$
and $-20~{\rm TeV}^{-2}<f_{WW}/\Lambda^2<10~{\rm TeV}^{-2}$.
These are still weaker than those obtained from $W^+W^+$ scattering at the
LHC \cite{ZKHY03} as given in Eq.~(\ref{LHCww}).
Therefore, $\gamma\gamma\to WWWW$ does not lead to an interesting process for
testing the anomalous gauge couplings of the Higgs boson at the 1 TeV ILC.

\begin{figure}[tb]
\includegraphics[width=16truecm,clip=true]{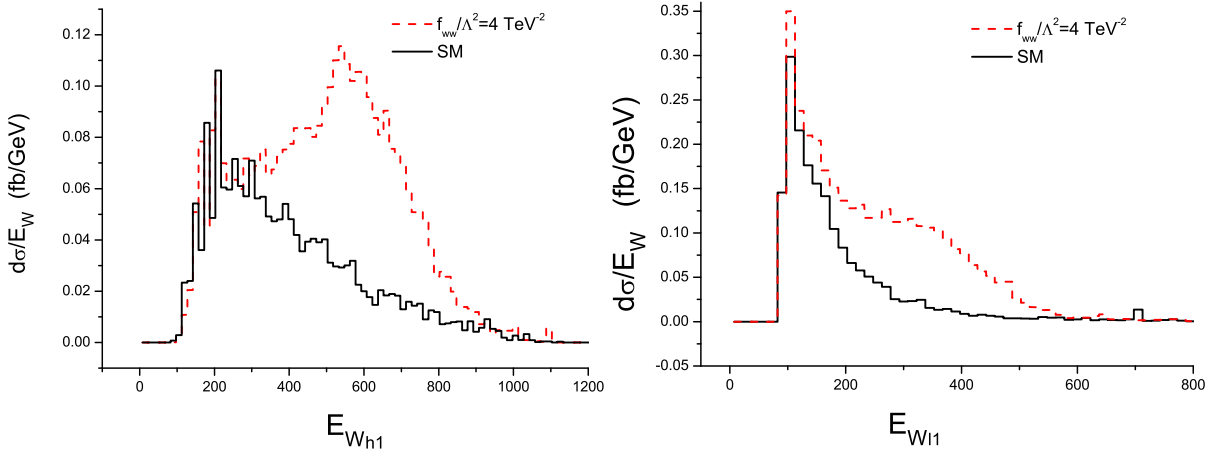}
\null\vspace{-0.2cm}
\caption{Energy distributions for $E_{W_{h1}}$ and $E_{W_{l1}}$
with $f_{WW}/\Lambda^2=4$ TeV$^{-2}$ and
$f_{WW}/\Lambda^2=0$ (SM).}
\label{e_small}
\end{figure}

\begin{figure}[tb]
\includegraphics[width=16truecm,clip=true]{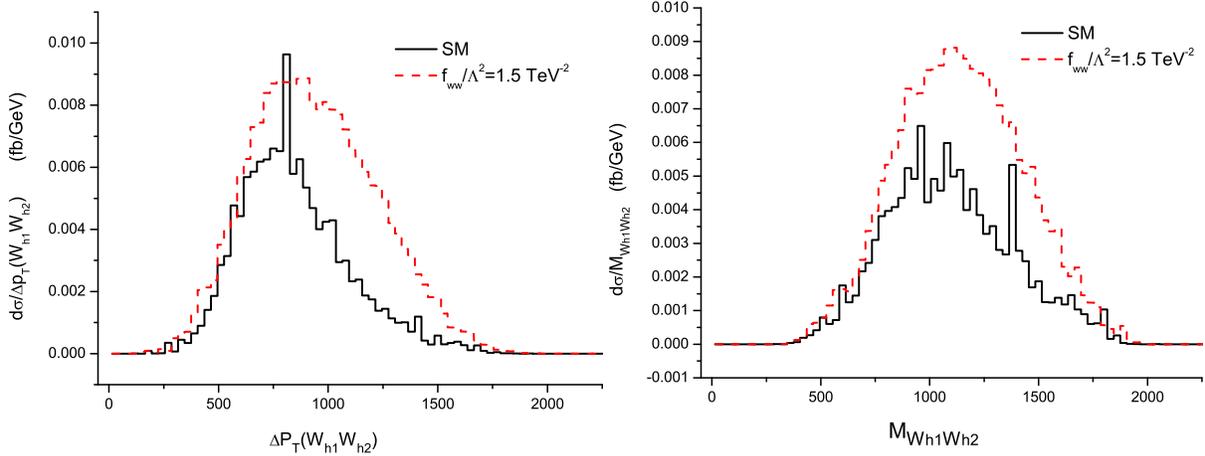}
\null\vspace{-0.2cm}
\caption{$\Delta P_T(W_{h1}W_{h2})$ and $M_{W_{h1}W_{h2}}$ distributions of the
$f_{WW}/\Lambda^2=1.5$ TeV$^{-2}$ and $f_{WW}/\Lambda^2=0$ (SM).}
\label{pt_mww}
\end{figure}

Obviously, one would like to consider a higher energy reach for a more favorable
kinematics for the 4-$W$ production. This leads us to look at the case of the 3 TeV CLIC.
Unlike the cases of the 500 GeV and 1 TeV ILC,
for the case of $f_{WW}/\Lambda^2$ dominance it is possible to impose certain kinematic cuts to
further improve the signal to background ratio effectively at this energy.
To see this, we divide the four final state $W$ bosons into two groups
according to their transverse momenta $P_T$. We denote the two $W$ bosons with higher $P_T$
by $W_{h1}$ and $W_{h2}$ with $P_T(W_{h1})<P_T(W_{h2})$,
for example the two scattered $W$ bosons in the central rapidity region as in FIG.~\ref{AA-WWWW}(a),
and denote those two with lower $P_T$ by $W_{l1}$ and $W_{l2}$ with
$P_T(W_{l1})<P_T(W_{l2})$, for example the two spectator $W$ bosons in FIG.~\ref{AA-WWWW}(a).
We first study the energy distributions $E_{W_{h1}}$ and $E_{W_{l1}}$ for
$f_{WW}/\Lambda^2=0$ (SM) and $f_{WW}/\Lambda^2\ne 0$ in FIG.~\ref{e_small}.
We see the differences between the $f_{WW}/\Lambda^2\ne 0$
and the SM background distributions.
The harder distributions for the signal motivate us to impose the following cuts
\begin{eqnarray}                       
E_{W_{h1}}>350~{\rm GeV},~~~~~~~~~~E_{W_{l1}}>200~{\rm GeV},
\label{E_small}
\end{eqnarray}
to effectively suppress the SM background.
We next introduce the transverse momentum difference
\begin{equation}
\Delta P_T(W_{h1}W_{h2})\equiv |\vec  P_T(W_{h1}) - \vec  P_T(W_{h2}) |^2 .
\end{equation}
In FIG.~\ref{pt_mww}, we plot this variable $\Delta P_T(W_{h1}W_{h2})$ and
the invariant mass  $M_{W_{h1}W_{h2}}$ of the higher $P_T$ $W$-pair for
$f_{WW}/\Lambda^2=1.5$ TeV$^{-2}$ and $f_{WW}/\Lambda^2=0$ (SM).
We see that the cuts
\begin{eqnarray}                            
\Delta P_T(W_{h1}W_{h2})>750~{\rm GeV},~~~~~~~~~~M_{W_{h1}W_{h2}}>850~{\rm GeV},
\label{pt-mww}
\end{eqnarray}
can help improve the signal to background ratio.

\begin{figure}[tb]
\includegraphics[width=16truecm,clip=true]{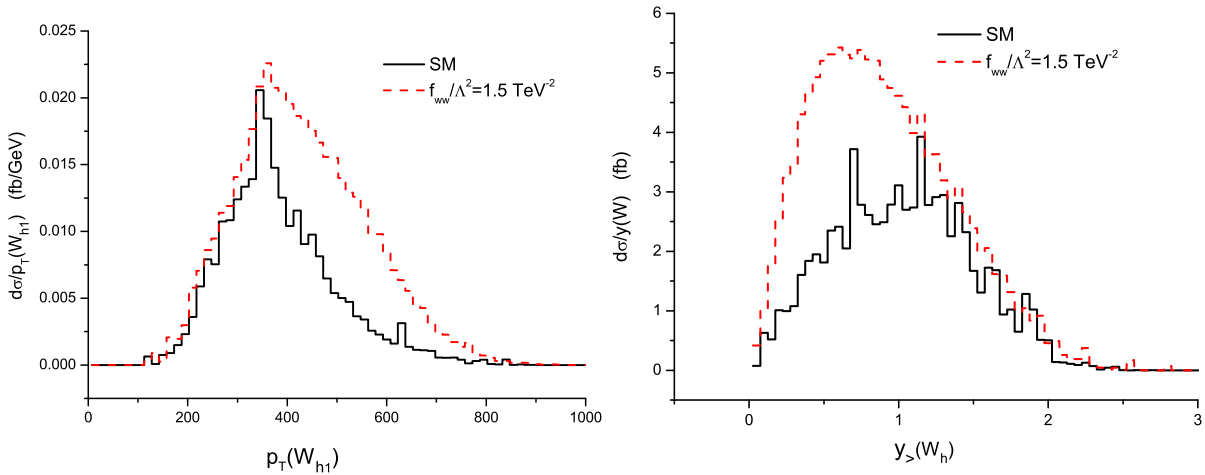}
\null\vspace{-0.2cm}
\caption{$P_T(W_{h1})$ and $y_>(W_{h})$ distributions of the
$f_{WW}/\Lambda^2=1.5$ TeV$^{-2}$ and $f_{WW}/\Lambda^2=0$ (SM).}
\label{pt_y}
\end{figure}

\begin{figure}[tb]
\includegraphics[width=16truecm,clip=true]{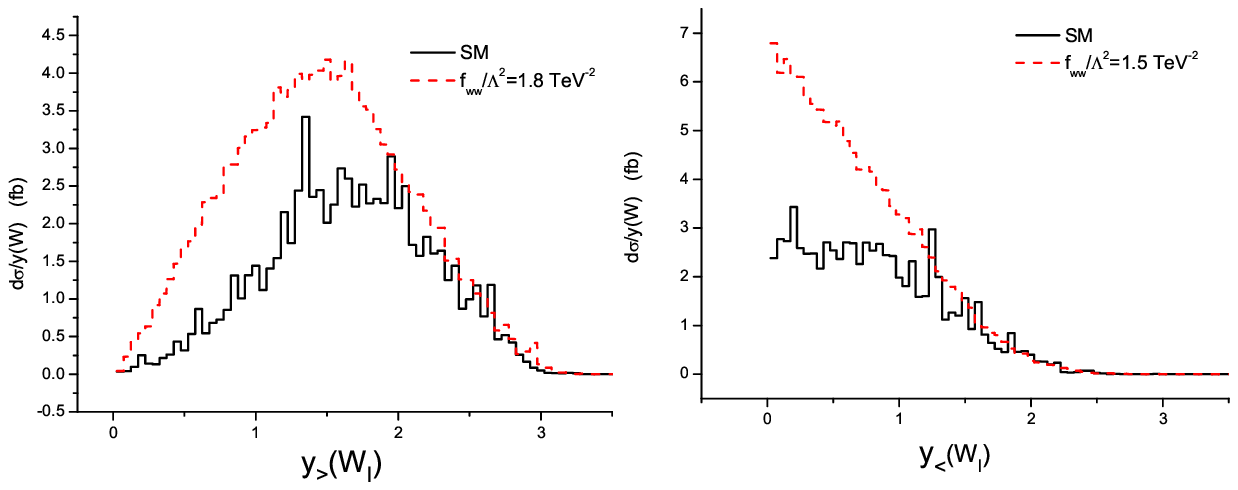}
\null\vspace{-0.2cm}
\caption{$y_>(W_l)$ and $y_<(W_l)$ distributions of the
$f_{WW}/\Lambda^2=1.5$ TeV$^{-2}$ and $f_{WW}/\Lambda^2=0$ (SM).}
\label{y+-y-}
\end{figure}

\begin{figure}[tb]
\includegraphics[width=16truecm,clip=true]{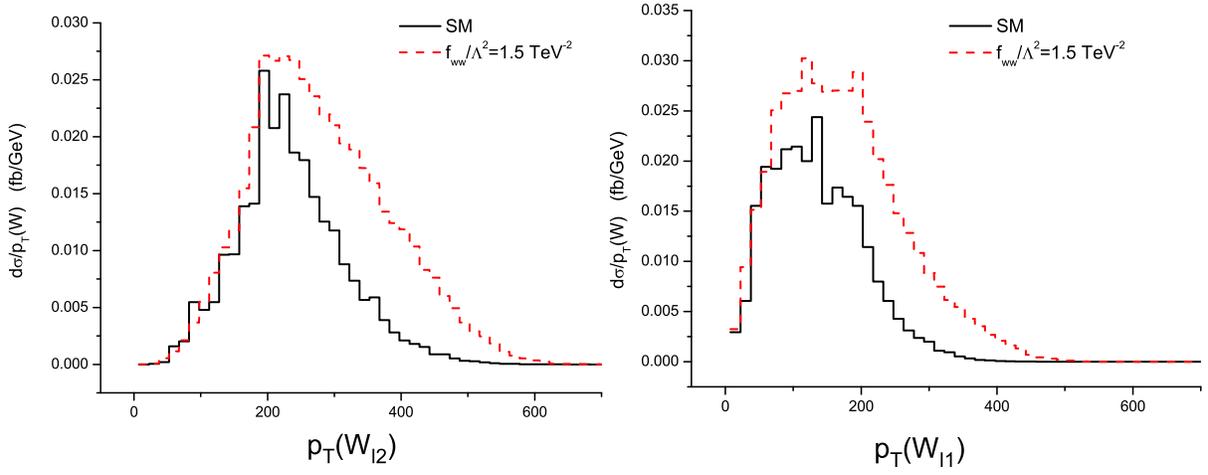}
\null\vspace{-0.2cm}
\caption{$P_T(W_{l2})$ and $P_T(W_{l1})$ distributions of the
$f_{WW}/\Lambda^2=1.5$ TeV$^{-2}$ and $f_{WW}/\Lambda^2=0$ (SM).}
\label{pt2pt1}
\end{figure}

We now examine the transverse momentum and the rapidity distributions.
Denote  the larger and smaller absolute rapidities
of $W_{h1}$ and $W_{h2}$ by $y_>(W_h)$ and $y_<(W_h)$,
respectively. FIGURE \ref{pt_y}
shows the $P_T(W_{h1})$ and $y_>(W_h)$ distributions for
$f_{WW}/\Lambda^2=1.5$ TeV$^{-2}$ and $f_{WW}/\Lambda^2=0$ (SM).
We see that the cuts
\begin{eqnarray}                          
p_T(W_{h1})>400~{\rm GeV},~~~~~~~~~~|y_>(W_h)|<1.2,
\label{pt-y}
\end{eqnarray}
can further improve the signal to background ratio.
FIGURE \ref{y+-y-} plots the $y_>(W_l)$ and
$y_<(W_l)$ distributions for the $f_{WW}/\Lambda^2=1.5$ TeV$^{-2}$ and
$f_{WW}/\Lambda^2=0$ (SM). We impose the following cuts
\begin{eqnarray}                            
{\rm both}~|y_>(W_l)|~{\rm and}~|y_<(W_l)|<2.0,\qquad |y_>(W_l)|~{\rm and/or}~|y_<(W_l)|<1.2.
\label{y>-y<}
\end{eqnarray}
Finally, we plot the $p_T(W_{l2})$ and $P_T(W_{l1})$ distributions for
$f_{WW}/\Lambda^2=1.5$ TeV$^{-2}$ and $f_{WW}/\Lambda^2=0$ (SM)
in FIG.~\ref{pt2pt1}, which suggest  the following optimal cuts
\begin{eqnarray}                        
{\rm both}~P_T(W_{l2})~{\rm and}~P_T(W_{l1})>100~{\rm GeV},
~~~~~~~~~~P_T(W_{l2})~{\rm and/or}~P_T(W_{l1})>250~{\rm GeV}.
\label{pt2-pt1}
\end{eqnarray}

\begin{table}[tb]
\caption{Detection efficiencies after various cuts for $f_{WW}/\Lambda^2=1.5$ TeV$^{-2}$ as an
example.}
\renewcommand{\baselinestretch}{1}
\tabcolsep 5pt
\begin{tabular}{ccccc}
\hline\hline
 &no cuts&cut (\ref{E_small})&cuts (\ref{E_small})+(\ref{pt-mww})+(\ref{pt-y})
 &cuts (\ref{E_small})+(\ref{pt-mww})+(\ref{pt-y})+(\ref{y>-y<})+(\ref{pt2-pt1})\\
\hline
$\sigma_{signal}$ (fb)&3.2&2.8&1.3&1.0\\
$\sigma_{bkgd}$~~~~(fb)&27&3.8&1.2&0.37\\
effectiveness $r$ &-&$87\%/14\%=6.2$&$46\%/31\%=1.5$&$78\%/31\%=2.5$\\
\hline\hline
\label{efficiency2}
\end{tabular}
\end{table}

\begin{figure}[tb]
\includegraphics[width=16truecm,clip=true]{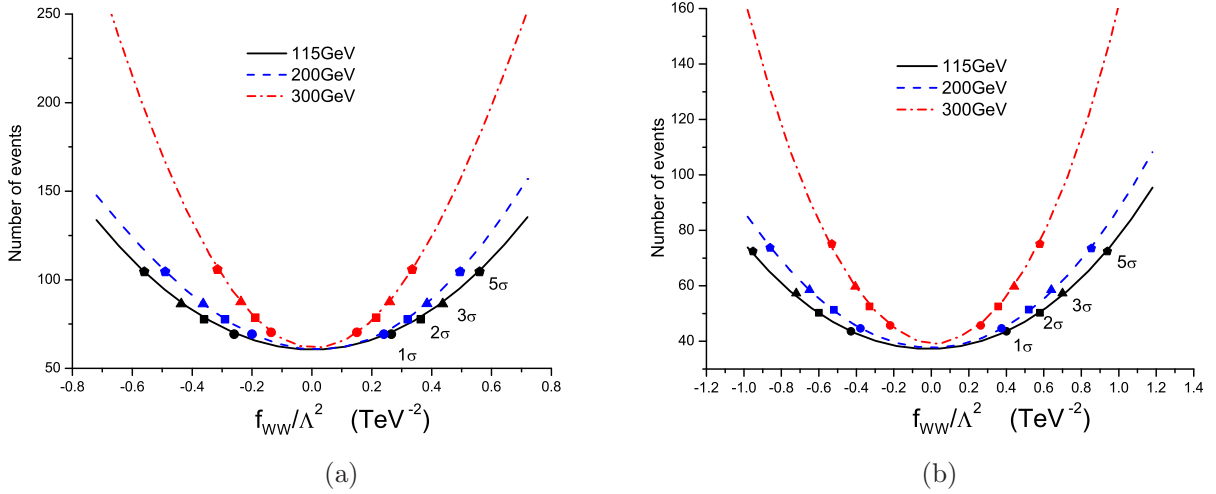}
\null\vspace{-0.5cm}
\caption{Number of events of $\gamma\gamma\to WWWW~(W\to jj)$
versus $f_{WW}/\Lambda^2$ for $m_H=115$, 200
and 300 GeV at the $\sqrt{s_{ee}}=3$ TeV CLIC
with the cuts (\ref{E_small})$\--$(\ref{pt2-pt1}) and taking account of
the $50\%$ safety factor for the final state $W$ boson detection:
(a) polarized, (b) unpolarized.
The values of $f_{WW}/\Lambda^2$ corresponding to
$1\sigma$, $2\sigma$, $3\sigma$ and $5\sigma$ statistical deviations
[cf.~Eq.~(\ref{N})] are shown on each curve by
the bullets, squares, triangles and asterisks, respectively.}
\label{AA4W-f_WW}
\end{figure}

With the imposed cuts (\ref{E_small}) $-$ (\ref{pt2-pt1}),
the SM background can be effectively suppressed.
The effectiveness of Eq.~(\ref{effi}) after imposing each of the cuts are shown in TABLE~\ref{efficiency2}.
Again, taking into account the $50\%$
safety factor in the final state $W$ boson reconstruction by $M_W$ from $M_Z$,
we obtain the number of events for $m_H=115,\ 130,\ 200$ and 300 GeV at the polarized
($2\lambda_eP_c=-1$) and unpolarized photon colloders based on the 3 TeV CLIC. The results
for the cases of $f_{WW}/\Lambda^2$ dominance are plotted in FIG.~\ref{AA4W-f_WW}.
The $1\sigma$, $2\sigma$, $3\sigma$ and $5\sigma$ statistical deviations $\sigma_{stat}$
are shown by the bullets, squares, triangles and asterisks, respectively,
along the $f_{WW}/\Lambda^2$ axis.
We see that, at the 3 TeV CLIC, the process $\gamma\gamma\to WWWW$ can sensitively
test $f_{WW}/\Lambda^2$.
The sensitivity for $m_H=115,~130$ and 200 GeV is similar, while that for
$m_H=300$ GeV
is significantly better just as in the case of $\gamma\gamma\to ZZ$.
From FIG.~\ref{AA4W-f_WW}
we see that the $2\sigma$ and $3\sigma$ sensitivities for $m_H=115,~130$ and 200 GeV are roughly
\begin{eqnarray}                          
&&\sqrt{s_{ee}}=3~{\rm TeV~polarized}~(m_H=115\--200~{\rm GeV}):\hspace{8cm}\nonumber\\
&&~~~~~~~~2\sigma:~~~~~~~~~~~~~~~~-0.4~{\rm TeV}^{-2}<f_{WW}/\Lambda^2<0.4~{\rm TeV}^{-2},\nonumber\\
&&~~~~~~~~3\sigma:~~~~~~~~~~~~~~~~-0.45~{\rm TeV}^{-2}<f_{WW}/\Lambda^2<0.45~{\rm TeV}^{-2},
\nonumber\\
&&\sqrt{s_{ee}}=3~{\rm TeV~unpolarized}~(m_H=115\--200~{\rm GeV}):\hspace{8cm}\nonumber\\
&&~~~~~~~~2\sigma:~~~~~~~~~~~~~~~~-0.6~{\rm TeV}^{-2}<f_{WW}/\Lambda^2<0.6~{\rm TeV}^{-2},
\nonumber\\
&&~~~~~~~~3\sigma:~~~~~~~~~~~~~~~~-0.8~{\rm TeV}^{-2}<f_{WW}/\Lambda^2<0.8~{\rm TeV}^{-2}.
\label{3000f_WW}
\end{eqnarray}
We note that although the  sensitivities in the polarized and unpolarized cases
are similar, they are  about a factor of 6 and a factor of 3.6 more sensitive than the
corresponding sensitivity $|f_{WW}/\Lambda^2|\le 2.9~{\rm TeV}^{-2}$ obtained
from $W^+W^+$ scattering at the LHC \cite{ZKHY03}.
One can translate the above bounds into
the corresponding sensitivities on $g^{(i)}_{HVV}$:
\begin{eqnarray}                
\null\noindent{\rm polarized}:&&\nonumber\\
2\sigma:&&-0.005~{\rm TeV}^{-1}<g_{H\gamma\gamma}<0.005~{\rm TeV}^{-1},\hspace{8cm}\nonumber\\
&&-0.011~{\rm TeV}^{-1}<g^{(1)}_{HWW}<0.011~{\rm TeV}^{-1},
-0.021~{\rm TeV}^{-1}<g^{(2)}_{HWW}<0.021~{\rm TeV}^{-1},\nonumber\\
&&-0.011~{\rm TeV}^{-1}<g^{(1)}_{HZZ}<0.011~{\rm TeV}^{-1},
~~-0.008~{\rm TeV}^{-1}<g^{(2)}_{HZZ}<0.008~{\rm TeV}^{-1},\nonumber\\
&&-0.006~{\rm TeV}^{-1}<g^{(1)}_{HZ\gamma}<0.006~{\rm TeV}^{-1},
~~-0.004~{\rm TeV}^{-1}<g^{(2)}_{HZ\gamma}<0.004~{\rm TeV}^{-1},\nonumber\\
3\sigma:&&-0.005~{\rm TeV}^{-1}<g_{H\gamma\gamma}<0.005~{\rm TeV}^{-1},\nonumber\\
&&-0.012~{\rm TeV}^{-1}<g^{(1)}_{HWW}<0.012~{\rm TeV}^{-1},
-0.023~{\rm TeV}^{-1}<g^{(2)}_{HWW}<0.023~{\rm TeV}^{-1},\nonumber\\
&&-0.012~{\rm TeV}^{-1}<g^{(1)}_{HZZ}<0.012~{\rm TeV}^{-1},
~~-0.009~{\rm TeV}^{-1}<g^{(2)}_{HZZ}<0.009~{\rm TeV}^{-1},\nonumber\\
&&-0.006~{\rm TeV}^{-1}<g^{(1)}_{HZ\gamma}<0.006~{\rm TeV}^{-1},
~~-0.005~{\rm TeV}^{-1}<g^{(2)}_{HZ\gamma}<0.005~{\rm TeV}^{-1}.
\label{3000gp'}
\end{eqnarray}
\begin{eqnarray}                
\null\noindent {\rm unpolarized}:&&\nonumber\\
2\sigma:~~&&-0.007~{\rm TeV}^{-1}<g_{H\gamma\gamma}<0.007~{\rm TeV}^{-1},\hspace{8cm}\nonumber\\
&&-0.016~{\rm TeV}^{-1}<g^{(1)}_{HWW}<0.016~{\rm TeV}^{-1},
-0.031~{\rm TeV}^{-1}<g^{(2)}_{HWW}<0.031~{\rm TeV}^{-1},\nonumber\\
&&-0.016~{\rm TeV}^{-1}<g^{(1)}_{HZZ}<0.016~{\rm TeV}^{-1},
~~-0.012~{\rm TeV}^{-1}<g^{(2)}_{HZZ}<0.012~{\rm TeV}^{-1},\nonumber\\
&&-0.008~{\rm TeV}^{-1}<g^{(1)}_{HZ\gamma}<0.008~{\rm TeV}^{-1},
~~-0.007~{\rm TeV}^{-1}<g^{(2)}_{HZ\gamma}<0.007~{\rm TeV}^{-1},\nonumber\\
3\sigma:~~&&-0.010~{\rm TeV}^{-1}<g_{H\gamma\gamma}<0.010~{\rm TeV}^{-1},\nonumber\\
&&-0.021~{\rm TeV}^{-1}<g^{(1)}_{HWW}<0.021~{\rm TeV}^{-1},
-0.042~{\rm TeV}^{-1}<g^{(2)}_{HWW}<0.042~{\rm TeV}^{-1},\nonumber\\
&&-0.021~{\rm TeV}^{-1}<g^{(1)}_{HZZ}<0.021~{\rm TeV}^{-1},
~~-0.016~{\rm TeV}^{-1}<g^{(2)}_{HZZ}<0.016~{\rm TeV}^{-1},\nonumber\\
&&-0.011~{\rm TeV}^{-1}<g^{(1)}_{HZ\gamma}<0.011~{\rm TeV}^{-1},
~~-0.008~{\rm TeV}^{-1}<g^{(2)}_{HZ\gamma}<0.008~{\rm TeV}^{-1}.
\label{3000gu'}
\end{eqnarray}

\noindent
The sensitivities in (\ref{3000gp'}) and (\ref{3000gu'}) are of the same order of magnitude but
somewhat weaker than that for $g_{H\gamma\gamma}$ in (\ref{3000Pfg}) and (\ref{3000Ufg})
as from the process $\gamma\gamma\to ZZ$. However,  the process
$\gamma\gamma\to WWWW$  contains information of couplings $g_{HVV}$'s
as seen in (\ref{3000gp'}) and (\ref{3000gu'}), thus the two processes are complementary.
Compared with the sensitivity in (\ref{3000gu'}),
Eq.~(\ref{3000gp'}) with polarized beams  improves the sensitivity by a factor of 1.5.

\begin{figure}[tb]
\includegraphics[width=16truecm,clip=true]{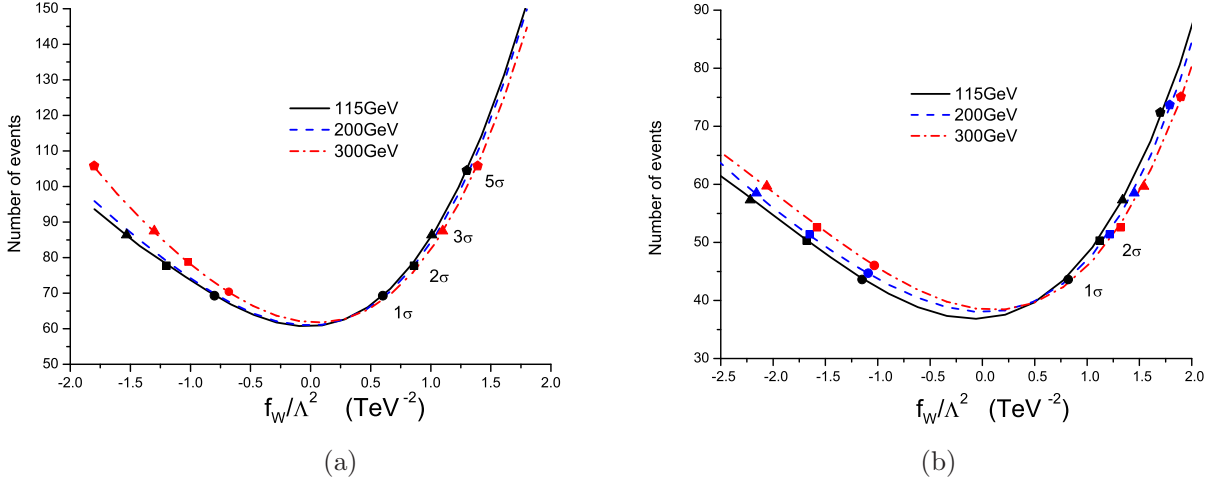}
\null\vspace{-0.5cm}
\caption{Number of events of $\gamma\gamma\to WWWW~(W\to jj)$
versus $f_W/\Lambda^2$ for $m_H=115$, 200
and 300 GeV at the $\sqrt{s_{ee}}=3$ TeV CLIC
with the cuts (\ref{E_small})$\--$(\ref{pt2-pt1}) and taking account of the $50\%$ safety factor
for the final state $W$ boson detection:
(a) polarized, (b) unpolarized.
The values of $f_W/\Lambda^2$ corresponding to $1\sigma$, $2\sigma$, $3\sigma$ and $5\sigma$ statistical deviations [cf.~Eq.~(\ref{N})] are shown on each curve by the
 bullets, squares, triangles and asterisks, respectively.}
\label{AA4W-f_W}
\end{figure}

Next we look at the case of $f_W/\Lambda^2$ dominance. As mentioned before, the contribution of
FIG.~\ref{AA-WWWW}(c) is negligible in this case, so that the signal rate is significantly smaller.
Adopting  the same cuts as given in Eqs.~(\ref{E_small})$-$(\ref{pt2-pt1}),
the results are plotted in FIG.~\ref{AA4W-f_W}.
We see that the asymmetry in the $f_W/\Lambda^2>0$ and $f_W/\Lambda^2<0$
regions due to the interference is more pronounced than that in the case of
$f_{WW}/\Lambda^2$ dominance (cf. FIG.~\ref{AA4W-f_WW}).
This is because that the SM background is more similar in size to the signal in the
this case. Based on FIG.~\ref{AA4W-f_W},
we obtain the $2\sigma$ and $3\sigma$ bounds
\begin{eqnarray}                              
{\rm polarized}:~~&&\hspace{10cm}\nonumber\\
2\sigma:&&~~~~~~~~-1.2~{\rm TeV}^{-2}\le f_W/\Lambda^2\le 0.8~{\rm TeV}^{-2},\nonumber\\
3\sigma:&&~~~~~~~~-1.6~{\rm TeV}^{-2}\le f_W/\Lambda^2\le 1.1~{\rm TeV}^{-2},\nonumber\\
\null\noindent {\rm unpolarized}:&&\hspace{10cm}\nonumber\\
2\sigma:&&~~~~~~~~-1.5~{\rm TeV}^{-2}\le f_W/\Lambda^2\le 1.2~{\rm TeV}^{-2},\nonumber\\
3\sigma:&&~~~~~~~~-2.4~{\rm TeV}^{-2}\le f_W/\Lambda^2\le 1.5~{\rm TeV}^{-2}.
\label{3000f_W}
\end{eqnarray}
Compared with the $3\sigma$ sensitivity
obtained from $W^+W^+$ scattering at the LHC \cite{ZKHY03},
the present result in the polarized case is slightly improved,
while that in the unpolarized case is a little less  sensitive.
As we have seen before, the sensitivity in the
polarized case is a factor of 1.5 better with respect to the unpolarized case.
The results again can be translated into sensitivities for $g^{(i)}_{HVV}$'s
\begin{eqnarray}                
\null\noindent {\rm polarized}:~~&&\nonumber\\
2\sigma:&&
-0.031~{\rm TeV}^{-1}<g^{(1)}_{HWW}<0.021~{\rm TeV}^{-1},
-0.042~{\rm TeV}^{-1}<g^{(2)}_{HWW}<0.062~{\rm TeV}^{-1},\hspace{0.5cm}\nonumber\\
&&-0.031~{\rm TeV}^{-1}<g^{(1)}_{HZZ}<0.021~{\rm TeV}^{-1},
~~-0.016~{\rm TeV}^{-1}<g^{(2)}_{HZZ}<0.024~{\rm TeV}^{-1},\nonumber\\
&&-0.017~{\rm TeV}^{-1}<g^{(1)}_{HZ\gamma}<0.011~{\rm TeV}^{-1},
~~-0.009~{\rm TeV}^{-1}<g^{(2)}_{HZ\gamma}<0.013~{\rm TeV}^{-1},\nonumber\\
3\sigma:&&
-0.042~{\rm TeV}^{-1}<g^{(1)}_{HWW}<0.029~{\rm TeV}^{-1},
-0.057~{\rm TeV}^{-1}<g^{(2)}_{HWW}<0.083~{\rm TeV}^{-1},\nonumber\\
&&-0.042~{\rm TeV}^{-1}<g^{(1)}_{HZZ}<0.029~{\rm TeV}^{-1},
~~-0.022~{\rm TeV}^{-1}<g^{(2)}_{HZZ}<0.032~{\rm TeV}^{-1},\nonumber\\
&&-0.022~{\rm TeV}^{-1}<g^{(1)}_{HZ\gamma}<0.015~{\rm TeV}^{-1},
~~-0.012~{\rm TeV}^{-1}<g^{(2)}_{HZ\gamma}<0.018~{\rm TeV}^{-1}.
\label{3000gp''}
\end{eqnarray}
\begin{eqnarray}                
\null\noindent {\rm unpolarized}:&&\nonumber\\
2\sigma:&&
-0.039~{\rm TeV}^{-1}<g^{(1)}_{HWW}<0.031~{\rm TeV}^{-1},
-0.062~{\rm TeV}^{-1}<g^{(2)}_{HWW}<0.078~{\rm TeV}^{-1},\hspace{0.5cm}\nonumber\\
&&-0.039~{\rm TeV}^{-1}<g^{(1)}_{HZZ}<0.031~{\rm TeV}^{-1},
~~-0.024~{\rm TeV}^{-1}<g^{(2)}_{HZZ}<0.030~{\rm TeV}^{-1},\nonumber\\
&&-0.021~{\rm TeV}^{-1}<g^{(1)}_{HZ\gamma}<0.017~{\rm TeV}^{-1},
~~-0.013~{\rm TeV}^{-1}<g^{(2)}_{HZ\gamma}<0.017~{\rm TeV}^{-1},\nonumber\\
3\sigma:&&
-0.062~{\rm TeV}^{-1}<g^{(1)}_{HWW}<0.039~{\rm TeV}^{-1},
-0.078~{\rm TeV}^{-1}<g^{(2)}_{HWW}<0.12~{\rm TeV}^{-1},\nonumber\\
&&-0.062~{\rm TeV}^{-1}<g^{(1)}_{HZZ}<0.039~{\rm TeV}^{-1},
~~-0.030~{\rm TeV}^{-1}<g^{(2)}_{HZZ}<0.048~{\rm TeV}^{-1},\nonumber\\
&&-0.034~{\rm TeV}^{-1}<g^{(1)}_{HZ\gamma}<0.021~{\rm TeV}^{-1},
~~-0.026~{\rm TeV}^{-1}<g^{(2)}_{HZ\gamma}<0.017~{\rm TeV}^{-1}.
\label{3000gu''}
\end{eqnarray}

In conclusion, the process $\gamma\gamma\to WWWW$ at the 3 TeV CLIC can provide a rather sensitive test of $g^{(i)}_{HVV}$ for the coupling $f_{WW}/\Lambda^2$, while  it only provides
a test of the coupling $f_W/\Lambda^2$ with  similar sensitivity to those obtained
from $W^+W^+$ scattering at the  LHC \cite{ZKHY03}.

\section{ SUMMARY}
\label{sum}

Once a light Higgs boson is discovered at the LHC and the ILC,
an immediate question is whether it is the SM-like or not.
Measuring the anomalous couplings of the Higgs boson may provide an
answer. So far, the most sensitive way of measuring the anomalous gauge couplings
of the Higgs boson at the LHC is via
$W^+W^+$ scattering \cite{ZKHY03}. In this paper, we have given a systematic study of the
sensitivity of testing the anomalous gauge couplings $g^{(i)}_{HVV}$'s of the Higgs boson via
$\gamma\gamma\to ZZ$ and $\gamma\gamma\to WWWW$
for a Higgs boson with $m_H=115\--300$ GeV at
polarized and unpolarized photon colliders based on $e^+e^-$ linear colliders of various energies.
We have developed certain kinematic cuts which can suppress the SM backgrounds effectively.

\begin{table}[tb]
\caption{Summary of the $3\sigma$ sensitivity for $g_{H\gamma\gamma}$ (in TeV$^{-1}$) from
$\gamma\gamma\to ZZ$ at various $e^+e^-$ linear colliders compared with the corresponding sensitivity  $-0.036<g_{H\gamma\gamma}<0.036$
from $W^+W^+\to W^+W^+$ at the LHC \cite{ZKHY03}.}
\vskip 1mm
\tabcolsep 8pt
\begin{tabular}{c}
\hline\hline
$3\sigma$ sensitivity for $g_{H\gamma\gamma}$ from $\gamma\gamma\to ZZ$ at the LC\\
\hline
500 GeV ILC ($m_H=115-200$ GeV): \\
\hspace{0.8cm}unpolarized: $-0.025<g_{H\gamma\gamma}<0.012$ \\
\hspace{0.8cm}polarized:~~~$-0.022<g_{H\gamma\gamma}<0.0082$ \\
\hline
1 TeV ILC ($m_H=115\--300$ GeV): \\
\hspace{0.8cm}unpolarized: $-0.0098<g_{H\gamma\gamma}<0.0073$\\
\hspace{0.8cm}polarized:$-0.0073<g_{H\gamma\gamma}<0.0054$ \\
\hline
3 TeV CLIC ($m_H=115\--300$ GeV):\\
\hspace{0.8cm}unpolarized: $-0.0022 < g_{H\gamma\gamma}<0.0022$ \\
\hspace{0.8cm}polarized:~$-0.0015<g_{H\gamma\gamma}<0.0015$  \\
\hline\hline
\end{tabular}
\label{S1}
\end{table}

The process $\gamma\gamma\to ZZ$ provides a sensitive test of the anomalous coupling $g_{H\gamma\gamma}$.
The kinematic cuts we proposed are (\ref{MZZ}), (\ref{theta_Z}) and (\ref{DeltaEcut}), which
effectively suppress the SM background. At the 500 GeV ILC, the obtained result shows that the
testing sensitivities for $m_H \le 2M_W$ are all similar, while that of $m_H=300$ GeV is much
higher due to the $s$-channel resonant production.
At the 1 TeV ILC and the 3 TeV CLIC, the $s$-channel
resonance effect is not so significant after imposing the cuts, and the sensitivities for
$m_H=115\--300$ GeV are all similar.
The obtained number of events and the statistic deviations are
shown in FIGs.~\ref{AAZZ500P}$\--$\ref{AAZZU}. We summarize the obtained testing sensitivities
[cf. (\ref{500Pfg}), (\ref{1000Pfg}), (\ref{3000Pfg}), (\ref{500Ufg}), (\ref{1000Ufg}), and
(\ref{3000Ufg})] in TABLE~\ref{S1}
together with the corresponding sensitivity obtained from $W^+W^+$ scattering at the LHC for
comparison.
For $m_H=300$ GeV at the 500 GeV ILC, the  $5\sigma$ sensitivity is
[cf.~Eq.~(\ref{500Pfg'})]
\begin{equation}
-0.0013~{\rm TeV}^{-1}<g_{H\gamma\gamma}<0.0085~{\rm TeV}^{-1}.
\end{equation}
If one can tune the energy of the ILC, it will be optimal
to make $E_{\gamma\gamma}$ peak at the resonant energy.
For instance, for $m_H=300$ GeV, we can tune the $e^+e^-$ energy to
$\sqrt{s_{ee}}=380$ GeV to have $E_{\gamma\gamma}$ to peak at 300 GeV.
In this case, the $5\sigma$ sensitivity is  [cf.~Eq.~(\ref{380Pfg})]
\begin{equation}
-0.00012~{\rm TeV}^{-1}<g_{H\gamma\gamma}<0.00030~{\rm TeV}^{-1}.
\end{equation}

\begin{table}[tb]
\caption{Summary of the $3\sigma$ testing sensitivities for $g^{(i)}_{HVV}$'s (in TeV$^{-1}$) for
$m_H=115\--200$ GeV from $\gamma\gamma\to WWWW$ at the 3 TeV CLIC compared with the corresponding sensitivity  from $W^+W^+\to W^+W^+$ at the LHC (from Ref.~\cite{ZKHY03}).}
\tabcolsep 2pt
\begin{tabular}{c|c}
\hline\hline
$3\sigma$ sensitivity for $g^{(i)}_{HVV}$ from $\gamma\gamma\to WWWW$ at the CLIC&
$3\sigma$ sensitivity for $g^{(i)}_{HVV}$ at the LHC\\
\hline
\null\hspace{-6cm}$f_{WW}/\Lambda^2$ dominant:&\null\hspace{-4cm}$f_{WW}/\Lambda^2$
dominant:\\
\hspace{-4cm}unpolarized: $|g_{H\gamma\gamma}|<0.010$,&\hspace{-1.9cm}$|g_{H\gamma\gamma}|<0.035$,\\
\hspace{0.55cm} $|g^{(1)}_{HWW}|<0.021,~|g^{(2)}_{HWW}|<0.042$&
\hspace{0.555cm} $|g^{(1)}_{HWW}|<0.075,~|g^{(2)}_{HWW}|<0.15$\\
\hspace{0.2cm} $|g^{(1)}_{HZZ}|<0.021,~|g^{(2)}_{HZZ}|<0.016$&
\hspace{0.35cm} $|g^{(1)}_{HZZ}|<0.075,~|g^{(2)}_{HZZ}|<0.058$\\
\hspace{0cm} $|g^{(1)}_{Hz\gamma}|<0.011,~|g^{(2)}_{HZ\gamma}|<0.008$&
\hspace{0.3cm}$|g^{(1)}_{Hz\gamma}|<0.041,~|g^{(2)}_{HZ\gamma}|<0.032$\\
\hspace{-4.2cm}polarized:~~~~$|g_{H\gamma\gamma}|<0.005$&\\
\hspace{0.55cm} $|g^{(1)}_{HWW}|<0.012,~|g^{(2)}_{HWW}|<0.023$&\\
\hspace{0.2cm} $|g^{(1)}_{HZZ}|<0.012,~|g^{(2)}_{HZZ}|<0.009$&\\
\hspace{0cm} $|g^{(1)}_{Hz\gamma}|<0.006,~|g^{(2)}_{HZ\gamma}|<0.005$&\\
\hline
\null\hspace{-6.2cm}$f_{W}/\Lambda^2$ dominant:&\null\hspace{-4.2cm}$f_{W}/\Lambda^2$ dominant:\\
\hspace{-4cm}unpolarized: ~~~~~~~~~~~~~~~~~~~~
&\\
\hspace{0.55cm} $-0.062<g^{(1)}_{HWW}<0.039,~-0.078<g^{(2)}_{HWW}<0.12$&
\hspace{0cm} $-0.047<g^{(1)}_{HWW}<0.042,\ -0.083<g^{(2)}_{HWW}<0.094$\\
\hspace{0.2cm} $-0.62<g^{(1)}_{HZZ}<0.039,\ -0.030<g^{(2)}_{HZZ}<0.048$&
\hspace{0cm} $-0.047<g^{(1)}_{HZZ}<0.042,\  -0.032<g^{(2)}_{HZZ}<0.036$\\
\hspace{0cm} $-0.034<g^{(1)}_{Hz\gamma}|<0.021,~-0.026<g^{(2)}_{HZ\gamma}|<0.017$&
\hspace{0cm} $-0.025<g^{(1)}_{Hz\gamma}<0.022,\ -0.018<g^{(2)}_{HZ\gamma}<0.020$\\
\hspace{-4.2cm}polarized:~~~~~~~~~~~~~~~~~~~~~
&\\
\hspace{0.55cm} $-0.042<g^{(1)}_{HWW}<0.029,~-0.057<g^{(2)}_{HWW}|<0.083$&\\
\hspace{0.2cm} $-0.042<g^{(1)}_{HZZ}<0.029,~-0.022<g^{(2)}_{HZZ}<0.032$&\\
\hspace{0cm} $-0.022<g^{(1)}_{Hz\gamma}<0.015,~-0.012<g^{(2)}_{HZ\gamma}<0.018$&\\
\hline\hline
\end{tabular}
\label{S2}
\end{table}

As for the process $\gamma\gamma\to WWWW$,
the 500 GeV and 1 TeV ILC cannot deliver large enough signal rates due to phase space
suppression. At the 3 TeV CLIC however, the $\gamma\gamma\to WWWW$ process
becomes interesting.
We have also developed certain kinematic cuts to suppress the SM background
[cf. (\ref{E_small})$\--$(\ref{pt2-pt1})]. The obtained number of events and the statistic deviations
are shown in FIGs.~\ref{AA4W-f_WW} and \ref{AA4W-f_W}.
We summarize the obtained $3\sigma$ testing
sensitivities [cf. (\ref{3000gp'}), (\ref{3000gu'}), (\ref{3000gp''}), and (\ref{3000gu''})] in
TABLE~\ref{S2} together with the corresponding sensitivities obtained from
$W^+W^+$ scattering at the LHC for comparison.

We have presented our results with an integrated luminosity of 1 ab$^{-1}$, and
the sensitivities are obtained by statistical errors only. We found that the beam
polarization is beneficial to the search. The polarized photon colliders can improve
the sensitivities by roughly a factor of 1.2$\--$1.4 for $\gamma\gamma\to ZZ$ and a factor of 1.5
for $\gamma\gamma\to WWWW$ relative to the unpolarized photon colliders.

In conclusion, photon colliders based on $e^+e^-$ linear colliders can provide more sensitive tests
of the anomalous gauge couplings of the Higgs boson than the LHC does. If a Higgs boson
candidate is found at the LHC, then linear colliders can give a better answer to whether
the Higgs boson is SM-like  or not. In the absence of new particle observation beyond a
light Higgs boson, testing anomalous Higgs boson couplings may
provide a way of discovering the effect of new physics beyond the SM.

\begin{center}
{\bf ACKNOWLEDGMENT}
\end{center}

This work is supported by National Natural Science Foundation of
China under Grant No. 90403017 (YPK and BZ) and No. 10435040 (BZ), and by U.S. DOE under Grant No.
DE-FG02-95ER40896 and Wisconsin Alumni Research Foundation (TH).


\end{document}